\documentclass[a4paper,11pt]{article}
\pdfoutput=1 % if your are submitting a pdflatex (i.e. if you have
\usepackage{jheppub} % for details on the use of the package, please
\usepackage[T1]{fontenc} % if needed

\usepackage{tikz}
\usetikzlibrary{arrows.meta, decorations.markings}

\title{\boldmath  Sine-Liouville gravity as a Vertex Model on Planar Graphs}

\author[a,b]{ Ivan
  Kostov }

 \affiliation[a]{Universit\'e Paris-Saclay, CNRS, CEA, Institut 
 th\'eorique \\
	 91191 Gif-sur-Yvette, France} \affiliation[b]{ Beijing Institute
	 of Mathematical Sciences and Applications \\
  Huairou 101408,Beijing,China
 }

\emailAdd{ivan.kostov@ipht.fr}

 \abstract{ We investigate the universal behaviour of a one-parameter
 generalisation of the six-vertex model on planar graphs, which we
 refer to as the seven-vertex model, or 7vM for quick reference.  The
 7vM is characterised by a temperature coupling and its continuum
 limit exhibits massive, dilute and dense phases similarly to the
 $O(n)$ loop model.  However, there is an important distinction: the
 loop weights are no longer topological and the dynamics of the loops
 is now entangled with the local geometry of the lattice.  From the
 dual matrix model we derive explicit expressions for the sphere and
 disk partition functions in the continuum limit.  The disk partition
 function for fixed length is a deformation of the Bessel integral
 known as the Kr\"atzel function.  We argue that the 7v matrix model
 (7vMM) and Matrix Quantum Mechanics (MQM) provide two complementary
 non-perturbative realisations of sine-Liouville gravity.
 Specifically, we find that the continuum limits of 7vMM and MQM share
 the same classical spectral curve but describe two different types of
 branes in sine-Liouville gravity.  The 7vMM precisely covers the
 range of parameters where the Minkowskian MQM lacks a simple
 interpretation in terms of multiple tachyon scattering.  We
 investigate the flow relating the dilute and dense phases and argue
 that this flow is the gravitational analogue of the massless flow in
 the sine-Gordon model with imaginary mass coupling.  The two
 endpoints of the flow are described by a free boson coupled to
 Liouville gravity and compactified on circles with two different
 radii.  \\
 \bigskip
 
 \noindent  In memory of Ivan Todorov }

   \def\O{\Omega}
  
  \newcommand\re[1]{({\ref{#1}})}
 \def\be{\begin{eqnarray} } \def\ee{\end{eqnarray}} 
    \def\CA{{\mathcal{A}}} 
     \def\CS{\mathcal{S}}
      \def\CO{{\mathcal{ O} }} 
     \def\CL{{ \mathcal{ L} }}
   \def\bee{\be\begin{aligned}}
   \def\eee{\end{aligned} \ee }
      
   \def\la{\label}
  \def\<{\langle\,} 
  \def\>{\,  \rangle} 
 \def\CF{{\cal F}} 
 
  \def\p{\partial} 
  \def\a{\alpha}
 \def\X{\mathbf{X}}

 \def\uv{{_{\text{UV}}}}
 \def\ir{{_{\text{IR}}}}
 
\def\hb{\hbar}

   \def\k{ \kappa}

   \def\bS{{\mathbf S}}
  \def\bU{{\mathbf U}}

  \usepackage{bm} 

\def\ee{\end{eqnarray}}  \def\IR{{\mathbb{R}}}
\def\IZ{{\mathbb{Z}}} \def\IC{{\mathbb{C}}} 
  \def\la{\label} \def\l {\lambda}
 
\def\({\left(} \def\){\right)}
 \def\[{\left[} \def\]{\right]}
  
\def\hf{ {\textstyle{1\over 2}} }
\def\CZ{{ \mathcal{ Z} }}
\def\L{\Lambda}
\def\o{\omega} 
\def\CC{ {\mathcal C}}
\def\CG{\mathcal{G}} 
 \def\p{\partial}
  \def\a{\alpha} 
  \def\b{\beta} 
  
\def\e{\epsilon} 

 \def\t{\tau} 
 \def\vp{\varphi}

\def\reg{{\infty }}
 \def\Tr{{\rm Tr}}
\def\d{\delta}
 \def\CM{\mathcal{M}}
\def\Xpm{ X_\pm }
\def\Xp{ X_+ }
\def\Xm{ X_- }
\def\bb{  {M _\mathrm{b}}}

 \def\bPhi{\mathbf{\Phi}}

\def\bS{\mathbf{S}}
 \def\bO{\mathbf{\Omega}}
\def\qq{q}\def\qqp{q^{1/2}}\def\qqm{q^{-1/2}}
 \def\RSLd{ R^{_{{SL}}}_{_{\rm{dense}}}}
\def\RSGc{ R^{_{{SG}}}_{_{\rm{UV}}}}
  \def\RSGd{ R^{_{{SG}}}_{_{ \rm{IR}}}}
\def\xb{x_{\mathrm{b}}}\def\sb{s_{\mathrm{b}}}
\def\mra{\mathrm{a}}\def\mrb{\mathrm{b}}
\def\RSLc{ R^{_{{SL}}}_{_{\rm{UV}}}}
 \def\RSLd{ R^{_{{SL}}}_{_{ \rm{IR}}}}
 \def\SL{{{SL}}}
 \def\SG{{{SG}}}
\def\tpm{{\mu_\pm}}
\def\tp{{\mu_+}}\def\tm{{\mu_-}}
   \def\bal{\bm{\alpha}}
  \def\bY{\mathbf{Y}}
\def\bH{\mathbf{H}}
\def\bJ{\mathbf{J}}

\def\bmu{{\tilde{\kappa}}}
   \def\CT{\mathcal{T}}
\def\Rsluv{R^{^{\mathrm{SL}}}_{^{\uv}}}
\def\Rslir{R^{^{\mathrm{SL}}}_{^{\ir}}}

\begin{document} 
\maketitle

\flushbottom

\section{Introduction and summary of the results}
\label{sec:intro}

Sine-Liouville gravity, or sine-Gordon model coupled to 2D gravity, is
a rich and fascinating theory.  Although it has been studied for more
than three decades, starting with \cite{Moore:1992ac,Hsu:1992cm}, some
aspects of the theory, especially concerning the boundary behaviour,
are still poorly understood.

 The sine-Liouville gravity is closely related to the sine-Liouville
 CFT which, according to the Fateev-Zamolodchikov-Zamolodchikov (FZZ)
 conjecture \cite{FZZconj, 2017} proved rigorously in
 \cite{Hikida_2009}, is mapped by strong/weak coupling duality to the
 Euclidean black hole (cigar) CFT
 \cite{PhysRevD.44.314,Dijkgraaf:1991ba}.  A non-perturbative
 description of the sine-Liouville gravity is provided by Matrix
 Quantum Mechanics (MQM)
 \cite{KAZAKOV1988171,Brezin:1989ss,ginsparg19902d,gross1990nonperturbativest}
 perturbed by a source of winding \cite{Kazakov:2000pm} or momentum
 \cite{Alexandrov:2002fh} modes.  The perturbed MQM exhibits Toda
 integrability, first discovered in \cite{Dijkgraaf:1993aa} and then
 exploited to investigate different aspects of the 2D string theory:
 winding mode \cite{Kazakov:2000pm} and momentum mode
 \cite{Alexandrov:2002fh,Alexandrov:2001cm} condensates, and
 non-perturbative effects due to sine-Liouville instantons
 \cite{Alexandrov:2005ac,2024JHEP...01..141A}.
    
\begin{figure}[h!]
        \centering
%         %%----start of first figure----
\begin{minipage}[t]{0.8\linewidth}
            \centering
             \includegraphics[width= 11 cm]{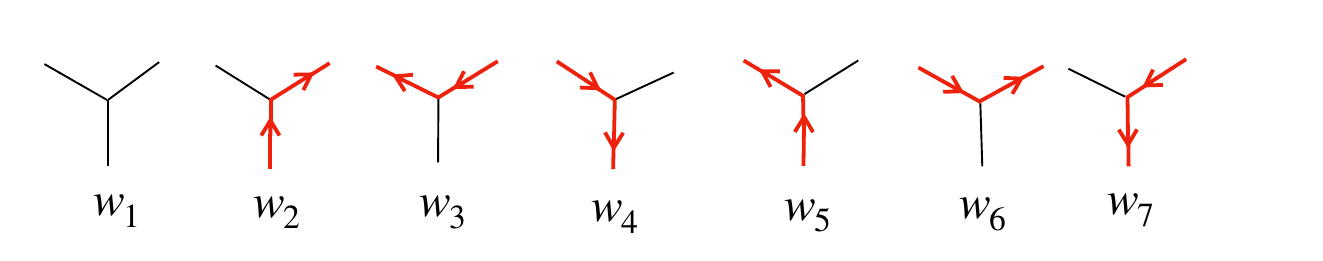}
 \caption{\small  The   vertices of the 7-vertex model.}
     \label{vertices}
          \end{minipage} 
\end{figure}

In this paper we present an alternative holographic dual of Euclidean
sine-Liouville gravity with a neat statistical interpretation.  Our
construction is based on the integrable lattice regularisation of the
sine-Gordon quantum field theory by the \emph{honeycomb seven-vertex
model}.\footnote{This statistical model has been given different names
in the literature.  We adopted the name ``seven-vertex model''
following Batchelor and Bl\"ote \cite{batchelor1989bethe}.  We
consider the formulation of the 7-vertex model in terms of arrows on a
honeycomb lattice \cite{sol7v} because it is most naturally
generalised to dynamical lattices.  } This vertex model, first studied
by Baxter \cite{baxter1987chromatic,baxter1986q}, gives a local
formulation of the $O(n)$ model \cite{Nienhuis:1984wm}.  The local
degrees of freedom of the vertex model are arrows assigned to the
bonds of the honeycomb lattice obeying the ``ice rule'' that the
numbers of incoming and outgoing arrows at each site must be equal.
The ice rule allows seven arrow configurations at each vertex of the
honeycomb lattice as shown in fig.  \ref{vertices}.  The local
Boltzmann weights are associated with these seven vertex
configurations, or simply vertices.  Assuming invariance under $\pi/3$
rotations, which we must do when we consider dynamical lattices, there
are only three distinct vertices which can be parametrised up to a
common factor as
\bee \la{7vflat} w_1= T, \quad w_2=w_3=w_4 = w , \quad w_5=w_6=w_7 =
w^{-1} .  \eee
The parameter $T $ is sometimes referred to as the temperature
coupling or shortly the temperature.  Each vertex configuration
defines a collection of self-avoiding and mutually avoiding oriented
loops, and the partition sum of the vertex model is expanded as
\bee
\la{loopexpansion}
Z = (T)^{\text{(sites)}} \sum_{\text{oriented loops}} (1/T)^{  
(\text{occupied sites}) } \ w^{(\text{left turns})-(\text{right
turns})}.
\eee
Since on the infinite lattice the number of the right turns differs
from the number of the left turns by $\pm 6$, the loop expansion
\re{loopexpansion} is that of the $O(n)$ model with $n= w^6+w^{-6}$
\cite{Nienhuis:1984wm}.  The 7v model is exactly solvable with the
special choice for the temperature coupling $T_\pm =(2 \pm
\sqrt{2-n})^{1/2}$
\cite{BLOTE1982405,baxter1987chromatic,baxter1986q,sol7v}.  The loop
gas exhibits nontrivial critical behavior when $w$ is a unimodular
complex number so that $|n|\leq 2$.  Its phase diagram shows two
critical phases, the phase of dilute loops at $T _c = T_+$ and the
critical phase of dense loops at $T<T_c$ which includes the other
solvable point $T_-$.  The two critical phases correspond to the two
renormalisation-group fixed points of the sine-Gordon theory with
\emph{purely imaginary} mass coupling and the massless flow connecting
them was investigated in a series of papers by P. Fendley, H. Saleur
and Al.  Zamolodchikov
\cite{Fendley:1993wq,Fendley:1993xa,Zamolodchikov:1994za}.

\begin{figure}[h!]
        \centering
%         %%----start of first figure----
\begin{minipage}[t]{0.7\linewidth}
            \centering
             \includegraphics[width= 11 cm]{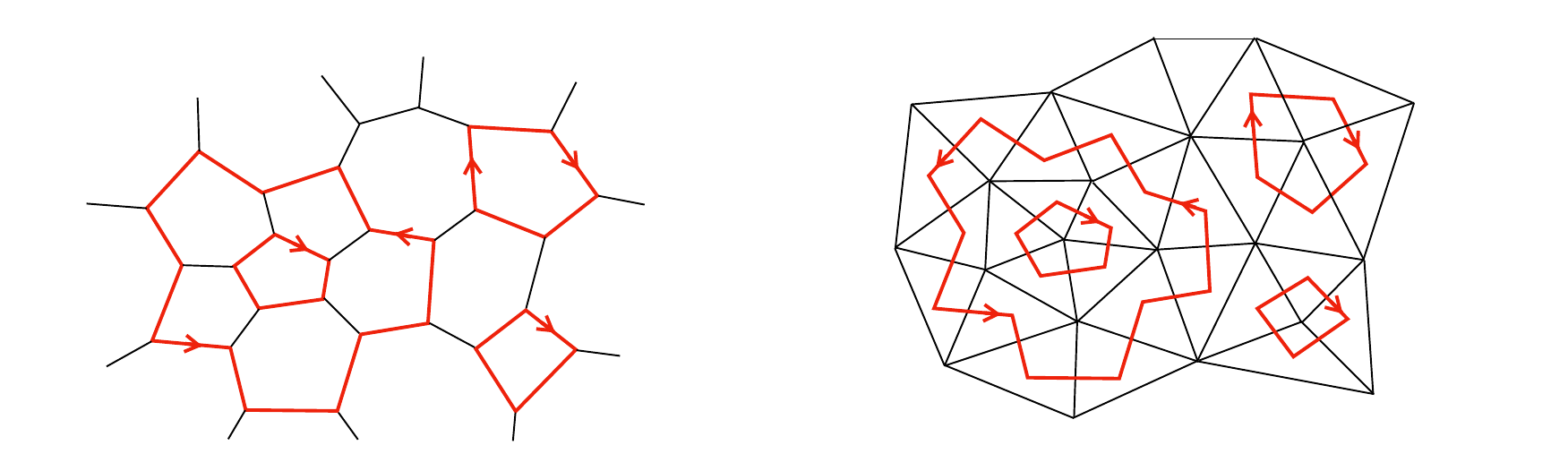}
\caption{\small A loop configuration on a trivalent graph with the
topology of the disk and on its dual triangulation.  To preserve the
information about the topology the lines should be thickened.  }
    \label{latticess}     \end{minipage} 
\end{figure}

 The 7-vertex model (7vM) has the advantage, compared to the other
 discretisations of the sine-Gordon model, that its weights are
 isotropic and therefore can be defined on any trivalent planar graph.
 Sometimes it is useful to represent the trivalent graph by its dual
 triangulation where we can define local curvature associated with the
 vertices of the triangulation.  An example of a vertex configuration
 on a trivalent graph with the topology of a disk together with its
 dual triangulation is shown in fig.  \ref{latticess}.  In the drawing
 the triangles had to be deformed but they are assumed equilateral.
 Considering the ensemble of all trivalent planar graphs, we add the
 gravitational degrees of freedom to achieve an integrable lattice
 regularisation of the gravitational sine-Gordon model.

The 7-vM on planar graphs can be reformulated as a gas of oriented
self-avoiding and mutually avoiding loops and in this respect
resembles the gravitational $O(n)$ loop model \cite{Kostov:1988fy}.
The dynamics is however quite different because the Boltzmann weights
of the loops depend on the local curvature through an effective spin
connection.  In general, vertex models on planar graphs represent a
separate and still unexplored niche of solvable models of 2D gravity,
and can reveal new unexpected phenomena, especially in the case of
lattices with boundaries.

We solve the gravitational 7vM by mapping it to a large-$N$ matrix
model representing a one-parameter deformation of the 6-vertex matrix
model formulated originally by Ginsparg \cite{Ginsparg:1991bi} and
solved in \cite{Kostov:1999qx,ZinnJustin:1999wt,ELVEYPRICE2023105739}.
We parametrise the vertices in fig.  \ref{vertices} in a way
compatible with the conventions of \cite{Kostov:1999qx} by setting in
\re{7vflat}
\bee
\la{7vflat1}
w=   e^{ i \pi \l /2}. 
\eee
Interesting critical behaviour exists only for real $\l$.  By the
symmetries of the Boltzmann weights one can restrict $0\le \l <1$.
For $\l=0$, the 7v matrix model is identical to the $O(2)$ matrix
model \cite{Kostov:1992pn}.

 In this paper, we are focusing on the universal behaviour in the
 scaling limit which is expected to be described by sine-Liouville
 gravity and which is characterised by two renormalised coupling
 constants, the cosmological constant $\mu$ and the temperature
 coupling $t$.  In order to keep the parallel with MQM, we work in the
 grand canonical ensemble where the cosmological constant $\mu$ is
 defined as a chemical potential for $N$, see e.g. the review
 \cite{1991hep.th....8019K}.  The grand partition function is a
 Fredholm determinant and can be viewed as an ensemble of free
 fermions with fixed Fermi energy $\mu$.  We formulate the Virasoro
 constraint on the spectral curve which is topologically an infinite
 (for $\l$ irrational) cover of the two-dimensional torus, with one
 periodic and one quasi-periodic cycles.  We solve the Virasoro
 constraint explicitly in the scaling limit where the periodic cycle
 becomes infinite.  The solution is unique up to renormalisations of
 the two coupling constants $\mu$ and $t$.  The exact normalisations
 in terms of the bare lattice parameters, which are not relevant for
 the discussion here, can be extracted from the full solution of the
 matrix model will be reported in a forthcoming paper
 \cite{Andre-Ivan-Mateus}.
 
We will use at different places the parameters $ \b$, $q$ and $ b $
related to the parameter $\l\in [0,1]$ by
\bee \la{parameterss} 
\boxed{\b = \pi (1+\l), \quad 
 q =- w^2=  e^{i\b}, \quad \textstyle b= \sqrt{1-\l\over 1+\l}}.
\eee
 With these conventions, the classical spectral curve in the scaling
 limit can be given the following parametric form, with $\l$
 considered as a continuous external parameter,
\bee
  \label{solscall}
x &= M (\o^b - \o^{ -1/b} ) , \\
 y &= M ^{ 1/b^2 } \, \o^{1/b} -t\, M ^{ b^2 } \o^{-b}.  \eee
Here the quantity $M $ is a function of $\mu$ and $t$ determined by
the (in general transcendental) equation
 \bee \la{MLpak00} M^{1+ \frac{1}{b^2}}- b^2 t M^{1+b^2} = (1+b^2)
 \mu.  \eee
The meromorphic function $y(x)$ defined by eq.  \re{solscall}
determines the partition function on the disk with complexified
boundary parameter $x$, and $M $ is the boundary entropy generated by
the fluctuations in the bulk of the disk.  The observables depend on
the couplings $\mu$ and $t$ and are generaly given, up to a power
factor depending on their dimension, by scaling functions of the
dimensionless variable $\hat t = t/\mu^{2\l/(1+\l)}$.

 We derive the thermodynamical equation of state for the
 susceptibility $u (\mu, t)= \p_\mu ^2\CF_0 $, where $\CF_0$ is the
 partition function on the sphere,
\bee\la{eqstate6v} \textstyle {\mu } = {1+\l\over 2} e^{ (1+\l)u } -
{1-\l \over 2}\, t \, e^{ (1-\l) u }.  \eee
 This equation is a consequence of \re{MLpak00} and the relation
 between the boundary entropy and the susceptibility
    \bee \la{Mofmu} M^2  = e^{-  {1-\l^2\over 4} u } . 
\eee

The equation of state \re{eqstate6v} has three critical points $t=0,
t\to -\infty$ and $t=t_c>0$ corresponding to the dilute, the dense and
the massive phases of the loop gas.  The phase diagram sine-Liouville
gravity extracted from the critical behaviour of the 7v model matches
the phase diagram of the sine-Gordon model on the plane where we
observe two nontrivial and one trivial fixed points of the
renormalisation group.  The role of the scale parameter is played in
sine-Liouville gravity by the cosmological constant $\mu$ which
determines the size of the two-dimensional universe.  The limit of
large $\mu$, or equivalently small $t$, describes the UV limit of the
matter field which is that of a free compactified boson.  When $t$
becomes large and negative, the effective cosmological constant
$-\mu/t$ diminishes and eq.  \re{eqstate6v} describes the physics at
large distances.  The second critical point $t\to -\infty$ thus
corresponds to the IR fixed point of the sine-Gordon model with
\emph{imaginary} mass coupling $\sim \sqrt{t}$.  The transcendental
equation \re{eqstate6v} is the counterpart in sine-Liouville gravity
of the massless flow in the sine-Gordon model with imaginary mass
coupling $\sim \sqrt{t}= \pm i \sqrt{|t|}$ studied in
\cite{Fendley:1993wq,Fendley:1993xa, Zamolodchikov:1994za}.  Finally,
the third critical point $t=t_c$ is that of pure gravity and
corresponds to the massive phase.
 
We find that at $t=0$ and $t\to -\infty$ the matter field is that of
a free boson compactified at radii respectively $\Rsluv $ and $\Rslir
$ related to the parameter $\l$ of the 7-vertex model as
   \bee
   \la{RcompSL} R^{^{\mathrm{SL}}}_{^{\uv}} = {1+\l\over 2} , \quad
   R^{^{\mathrm{SL}}}_{^{\ir}} = {1-\l\over 2} .
  \eee
In our conventions, the self-dual radius is 1.

As expected, the equation of state \re{eqstate6v} coincides with that
of MQM \cite{Kazakov:2000pm, Alexandrov:2002fh,Kostov:2002tk}.  Thus,
restricting to the bulk observables, we have two different
non-perturbative realisations of the same worldsheet theory.  The
realisation proposed here unveils a relation between the phase
diagrams of the sine-Gordon before and after coupling to gravity which
has not been noticed in the previous works based on MQM.
 
On the other hand, the boundary observables in the two realisations of
sine-Liouville gravity are {\it different} because the two models
correspond to two different Riemann surfaces sharing the same spectral
curve.  In MQM, the boundary observables for a given boundary length
are expressed, at least for $t=0$ and for $t\to -\infty$, in terms of
Bessel-$K$ functions.  On the other hand, in 7vMM the boundary
observables are expressed in terms of a generalised Bessel integral
also known as the Kr\"atzel function \cite{kratzel1975integral}.

The plan of the paper is as follows.  We start by reviewing in sect.
\ref{sect:2} the phase structure (the RG fixed points and the flows
connecting them) of the sine-Gordon model, in particular the massless
flow for imaginary mass coupling.  Then we formulate the expected
gravitational analogue of the massless flow in terms of the worldsheet
description of sine-Liouville gravity.  In sect.  \ref{sect:3} we
define the partition function of the 7vM and derive its
representations in terms of a gas of oriented loops on planar graphs
and as an interaction-round-a-triangle height model.  Then we
formulate the dual large-$N$ matrix model (the 7vMM) and derive the
Dyson-Schwinger equations as a Virasoro constraint for a collective
bosonic field.  In sect.  \ref{sect:Therm} we formulate the classical
Virasoro constraint as a boundary value problem, derive the solution
at the critical point and in the continuum limit, and compute the
partition functions on the sphere and on the disk.  For the latter we
give analytic expressions both for fixed boundary parameter and for
fixed boundary length.  In sect.  \ref{section:7vMMMQM} we explain how
the spectral curve of 7vMM compares to that of MQM and the
consequences of the new understanding of the phase diagram for the
proposal of \cite{Kazakov:2000pm} for an MQM-based description of the
Euclidean ``cigar'' background.

 \section{Sine-Liouville gravity }
 \la{sect:2}
 
 In this section we present some general arguments in support of our
 claim that sine-Liouville gravity describes the continuum limit of
 the 7v model on planar graphs.  The principal statement here is a
 conjecture about the phase diagram of the theory to be verified by
 the exact solution of the 7v matrix model.  We start with reviewing
 the Coulomb gas description for the vertex model on a regular lattice
 which leads to the sine-Gordon model and sketch the phase diagram of
 the latter.

\subsection{The 7v model on a flat lattice and the sine-Gordon QFT  }

The 7-vertex model is mapped onto a height model with adjacent heights
equal or different by $\pi $ and renormalises at large distances to a
Coulomb gas \cite{Nienhuis:1984wm}.  The critical temperature $T=T_c$
corresponds to the trivial critical point of the Coulomb gas with all
charge fugacities equal to zero represented by a Gaussian field $\vp$
with Coulomb coupling $g= 1+3\l$ in our notation for the vertex
weights and identification $\vp=\vp+ 2\pi$.  The symmetry $\vp\to
-\vp$ turns the target space into an orbifold with twice smaller
radius $\pi$, see e.g. \cite{Ginsparg:1987eb}.

By rescaling of the gaussian field $\vp\to \vp /R$ with $R^2=g$, the
Coulomb coupling can be set to one.  To stress that the radius $R$
characterises the UV limit, sometimes we will denote $R=R_\uv$.  The
gaussian field is then defined by the action
 \bee 
 \la{actgbSG} \CA_{_{\text{FB}}} [\vp  ]&= {1\over 4\pi}  \int(\p_\mu
\vp )^2 
\la{Agauss}
\eee
and taking values in the circle of radius $R$, i.e. $\vp + 2\pi R
\equiv \vp $.  The basic operators are electric or vertex operators
$\mathcal{O}_{n}=e^{in \varphi/R}$ and magnetic or vortex operators
$\tilde{\mathcal{O}}_{m}=e^{i m R \tilde\vp}$ creating discontinuity
$2\pi mR $, with scaling dimensions respectively
  \bee\la{dimsem} x_{\pm n}= \hf (n/R)^2 \qquad \text{and}\qquad
  \tilde x_{\pm m}=\hf (mR)^2 \eee
 with $n$ and $m$  integers.
In our normalisation,
%\footnote{The  other widely  used convention 
%   is $r= R/\sqrt{2}  $
%   where the self-dual radius is $r_{\text{s.d}}= 1/\sqrt{2}$.}
 the electric-magnetic duality, or T-duality,  
maps 
\bee
n\leftrightarrow m, \qquad R\leftrightarrow 1/R.
\eee

In the 7v model there are no magnetic operators because of the ice
rule.  The perturbation with $t\sim T-T_c$ is triggered by a pair of
electric operators with charges $n=\pm 2$ \cite{Nienhuis:1984wm} which
are relevant for $1<R<2$.  In terms of the fugacities $\mu_{\pm 2} $
for these charges,
 \bee \la{tmumu} t= \mu_{2}\, \mu_{-2}.  \eee

When $ t>0$, the fugacities $\mu_{\pm 2} $ grow under renormalisation,
the Coulomb gas coupling $g=R^2$ increases thus making these
fugacities more relevant and the system flows to a trivial massive
phase.  If, however, $ t<0$, Nienhuis \cite{Nienhuis:1984wm} showed
that the Coulomb coupling constant decreases and eventually renders
the fugacities irrelevant.  As a consequence, the renormalised $\mu_2$
and $\mu_{-2}$ begin to diminish in magnitude and the model flows
towards another gaussian field with smaller compactification radius $
R_\ir <1$.

The local quantum field theory for this Coulomb gas is the sine-Gordon
QFT described in the UV limit by the action\footnote{The connection
with the traditional notation is $ \b_{ \mathrm{SG}} = \sqrt{8\pi}/R
$.}
\bee \la{actgbSG} \CA^{\text{orb}}_{_{SG}} [\vp ]&= \CA _{_{FB}} -
\mu_{2} \, e^{ 2i \vp/R} - \mu_{-2} \, e^{ -2 i\vp/R} \\
&=   \CA _{_{FB}} - \mu_{_{SG}} \int d^2 x \cos\( 2   \vp/R \) .
\eee
Because of charge conservation, the bulk observables depend only on
the product \re{tmumu}.  For $t>0$ which corresponds to real mass
coupling $ \mu_{_{SG}} $ the theory flows to a massive phase.  The
non-unitary thermal flow for $t<$ which corresponds to \emph{purely
imaginary} mass coupling $\mu_{_{SG}} $ was thoroughly studied by
Fendley, Saleur and Al.  Zamolodchikov
\cite{Fendley:1993wq,Fendley:1993xa,Zamolodchikov:1994za}.

In the UV, the dimensions of the perturbing operators $e^{\pm 2 i
\vp/R}$ are
\be x_{\pm 2 }= {2\over R^2}= 2\ {p \over p+1 } \ee
where we introduced another commonly used, and in particular in
\cite{Zamolodchikov:1994za}, parameter
  \bee \la{gofp}  p={1\over  R^2-1} . \eee
For small $t$, the SG perturbation can be considered as triggered by a
neutral local ``thermal'' operator with coupling constant $t$ and
dimension
	\bee \la{simSG} x_{_{\SG}} = x_2+x_{-2} -2 = 2 \, {p-1\over p+1} .
	\eee
One-loop RG calculations \cite{PhysRevB.16.1217,Nienhuis:1984wm}
support the following relation between the compactification radii of
the free boson at the two extremities of the flow, $
R=R^{^{\mathrm{SG}}}_{^{\uv}} $ at $t=0$ and
$R=R^{^{\mathrm{SG}}}_{^{\ir}} $ at $t\to -\infty$,
\bee (R^{^{\mathrm{SG}}}_{^{\uv}} )^2+ (R^{^{\mathrm{SG}}}_{^{\ir}}
)^2 = 2.  \la{RuvRir} \eee
 or in terms of the parameter $p$,
 \bee \la{R0e0A} 
 R^{^{\mathrm{SG}}}_{^{\uv}} =  
 \sqrt{1+ 1/p} ,
 \quad R^{^{\mathrm{SG}}}_{^{\ir}} = 
 \sqrt{1-1/p}.  \eee
The limit $p\to\infty$ corresponds to the BKT radius $R_{\text{BKT}}
=1$ which in this case is also the self-dual radius.  One can think of
the oriented loops as kinks interpolating between adjacent classical
vacua of the sine-Gordon potential.  In the massive phase $T>T_c$
($t>0$), the theory falls into one of the classical vacua.  Droplets
of other classical vacua can appear but they are exponentially
suppressed.  At $T=T_c$ ($t=0$), the kinks become massless and stay
massless for $T<T_c$ ($t<0$).  For large negative $t$ the kinks fill
densely the lattice and form a critical phase.

\subsection{The 7v model on a dynamical  lattice and the sine-Liouville QFT  }

\la{sect2:2}

The continuum limit of the 7v model on a dynamical lattice is expected
to be described by the gravitational sine-Gordon model.  A subtle
point here is the identification of the Coulomb gas coupling or
equivalently the compactification radius $R$.

For large negative temperature, the 7vM on dynamical triangulations is
equivalent to the 6v model on dynamical quadrangulations
\cite{Kostov:1999qx}.  From the mapping (valid only in the scaling
limit) of the dual matrix model to MQM it was concluded
\cite{Kostov:1999qx} that the continuum limit is described by a free
boson coupled to Liouville gravity and compactified at radius $R=
\hf(1-\l)\in [0,\hf]$.  The BKT radius is therefore twice smaller than
on the flat lattice,
\be
\la{RBKTD}
R_{\mathrm{BKT}} = \hf \qquad \text{(dynamical lattice)}.
\ee
This can be explained with the fact that while on the flat lattice the
phase factors of the loops are $w^{\pm 6}$ or $w^{\pm 4}$ depending on
the model, on the dynamical lattice they can be any integer power of
$w$.  Therefore all spectrum of electric charges \re{dimsem} is
present.\footnote{ It is known that the curvature defects change the
phase factors of the loops surrounding them in the same way the
electric charges do.  The consequences of the curvature defects for
the loop gas have been elucidated by O.~Foda and B.~Nienhuis
\cite{Foda:1988in}.} Hence the Coulomb gas for the loop gas on
dynamical lattice will renormalise according to the lowest charges
$n=\pm 1$ with dimension
  \bee\la{dimsem} x_{\pm 1}= \hf R ^{-2} \eee
 which implies that the BKT radius is  given by \re{RBKTD}.

A candidate for the worldsheet description of the continuum limit of
the 7MM is the theory of 2D gravity where the Euclidean sine-Gordon
field plays the role of a matter field, sometimes referred to as
sine-Liouville theory.  We prefer to use more exact term
sine-Liouville gravity.

This sine-Liouville gravity, also referred to as the gravitational
sine-Gordon model, was first studied by G. Moore \cite{Moore:1992ac}.
The sine-Liouville action has been discussed in different context in
the paper \cite{SEDRAKYAN1993256}.  In conformal gauge for the
gravitational field, the UV action depends on the sine-Gordon field
$\vp $ and on the scale factor $\phi$ of the metric known as Liouville
field.  Given a background metric $\hat g_{ab}$ with curvature $\hat
R$, the UV action is\footnote{This action is to be distinguished from
the action of the sine-Liouville conformal field theory, there is no
ghost sector, no Liouville potential, the marginal sine-Liouville
interaction is $e^{\a\phi} \cos (a\vp)$ with $a^2-\a^2 =2$, and the
Liouville background charge is $Q= 1/\a$.  If we ignore the ghost
sector, then the sine-Liouville CFT and sine-Liouville QG intersect at
$R=2/3$ and $\mu=0$.  The FZZ conjecture \cite{FZZconj,2017} states
that the sine-Liouville CFT is related by strong/weak coupling duality
to the ``cigar'' CFT with central charge ${3k\over k-2}-1$
\cite{PhysRevD.44.314,Dijkgraaf:1991ba}.  The mapping of the
parameters is $a=\sqrt{k}, \ \ \a = \sqrt{k-2}$.  }
 \bee \la{actionSG} \CA_{_\SL} [\phi, \vp ]&={1\over 4\pi} \int
 \limits_\CM d^2 z \sqrt{\hat g} \[ (\hat \nabla \phi)^2 + (\hat
 \nabla \vp)^2+
 2 \phi  \hat R \]
+ {\rm ghosts} \\
& +\mu \, \int d^2 z \sqrt{\hat g} \, e^{ 2 \phi } +\sum_{\pm} \tpm
\int d^2 z \sqrt{\hat g} \ e^{(2-{1\over R}) \phi} e^{\pm{ i\over R}
\vp } .  \eee
The exponents in the action are chosen so that the exponential
operators are strictly marginal.  Again, the path integral depends
only on the product $ t= \tp\tm$.  A shift $\phi\to \phi + \log (r^2)$
is compensated by $\mu\to\mu /r^2$ and $\tpm \to \tpm /r^\a$ with $\a=
2- 1/R$, so that the thermal coupling $t=\tp\tm$ scales as
  \bee \la{scalingSL} t\sim \mu^{ 2- 1/R} .  \eee
The phase diagram of sine-Liouville gravity is expected to be
qualitatively the same as that of the sine-Gordon model in the plane.
In particular, the massless flow in the sine-Gordon with imaginary
mass coupling should have its counterpart in the sine-Liouville
gravity.  We will refer to it as \emph{gravitational massless flow}.

The gravitational massless flow connects two theories of Liouville
gravity with compactified boson as matter field.  The parameter $R$ in
the UV action \re{actionSG} is the compactification radius on the UV
side, $ \RSLc\equiv R$.  The matter boson on the IR side has another
compactification radius $ \RSLd $.  We will argue in section
\ref{section:Radii}, based on the solution of the matrix model, that
\bee
\la{SLcSLd}
\RSLd=1-\RSLc.
\eee
Another question one can ask is whether the compactification radius of
the free boson is modified by switching on gravity and if so, what is
the precise relation.  In other words, is it possible to find a map
$f$ such that $ \RSLc= f( \RSGc)$ and $ \RSLd= f( \RSGd)$.  Let us
assume that such a map exists and it is the same for the UV and the IR
CFTs.  Comparing \re{SLcSLd} and \re{RuvRir} we conclude that $f(x) =
\hf x^2+ $ constant.  The constant should be zero because at the BKT
radius $R_\ir=R_\uv =1/2$.  Thus we arrive at the relation
 \bee   \RSLc= \hf ( \RSGc)^2 , \quad  \RSLd= \hf ( \RSGd)^2 .  \la{RSLRSG}
 \eee

 \section{The 7-vertex model on planar graphs}
   \la{sect:3}

  \subsection{Definition and loop expansion }
 \la{sect:7vmWS}

Starting with the discrete realisation of the sine-Gordon model by the
7v model, we construct, following
\cite{David:1984tx,Ambjorn:1985az,Boulatov:1986jd}, a discrete
integrable realisation of sine-Liouville gravity by considering the 7v
model in the ensemble of trivalent planar graphs.

Given the trivalent planar graph $\CG$, the 7vM degrees of freedom are
introduced by assigning orientations to some of the bonds of $\CG$ in
such a way that the number of the incoming and the outgoing arrows at
each vertex is zero.  Any such configuration gives rise to a set of
loops on $\CG$, and the partition function of the gas of loops takes
the same form as \re{loopexpansion}.
 
The partition function of the gravitational 7vM on the sphere is
then defined as a sum over all trivalent genus-zero planar graphs
$\CG$.
 \bee \CF_{0} (\k, T)&=\sum_{\text{ graphs\ } \CG} \kappa^{V_\CG} \
 Z_{\CG} (\k, T) \\
 Z_{\CG} (\k, T)
&=  \sum_{\text{loops}\ \CL \text{ on \ } \CG} (1/T)^{
 V_{\text{loops}}}\ \prod_{\CL} 
  w^{ (\text{left turns})-(\text{right
turns})}  , 
 \la{loopexp7v} 
\eee
where $w= \exp( \pi i \l/2)$.  Besides the temperature $T$ coupled to
the volume occupied by loops, $V_{\text{loops}}$, we introduce a new
parameter $\k$, the ``discrete cosmological constant'', coupled to the
total volume of the graph $ V_\CG$.  We will refer to the statistical
model defined by \re{loopexp7v} as the seven-vertex model on dynamical
lattice or simply as the gravitational 7v model.  The above definition
is generalised in an obvious way to the case of planar graphs with
boundaries and of higher genus.  For graphs with a boundary, we have
to introduce also a ``discrete boundary cosmological constant''
coupled to the boundary length.
         
The principal distinction between the loop expansions of the 7v model
on a flat and on a dynamical lattice is that in the second case the
fugacities of the loops are no longer topological: the dynamics of
the loops is now entangled with the local geometry of the lattice.
  
An alternative formulation of the gravitational 7v model is as an SOS
type {\it height model} on dynamical triangular lattices.  We can
think of any planar graph $\CG$ as the dual object to a triangulation
$\CT$ of the sphere or of the disk as in fig.  \ref{latticess}, right.
Each vertex configuration on $\CG$ determines, up to a constant, an
integer-valued height function, with heights assigned to the sites of
the dual triangulation $\CT$.  The oriented loops divide the planar
graph into domains of constant heights, with the oriented loops
appearing as domain boundaries.  The difference of the heights on both
sides of an oriented loop is $\pm 1$ depending on the orientation.
For a given loop configuration, all heights but one can be
reconstructed by taking into account the loop orientations.  The
Boltzmann weight of a height configuration on the triangulation $\CT$
is a product of local factors associated with the elementary triangles
$\Delta_{ijk} $,
 \bee \la{BweightsSOS} W_{\Delta} (h_1,h_2,h_3)&= \delta_{h_1h_2}
 \delta_{h_2h_3}\delta_{h_3h_1} +{1\over T} \delta_{h_1h_2} h_{h_2h_3}
 A_{h_3h_1} \, w^ { h_3-h_1 } \ +{\rm cyclic }, \\
A_{ab}&\equiv  \d_{a, b+1}+\d_{a, b-1} .  
\eee
The partition sum in \re{loopexp7v} is generated by expanding the
product of the Boltzmann weights in monomials and performing the sum
over the heights.  Discrete models of 2D gravity based on height
models on dynamical triangulations have been studied e.g. in
\cite{Kostov:2004pt}.

The weights of the loops in the expansion \re{loopexp7v} can be given
a geometrical description in terms of the local curvature of the
discretised world surface.  For that we must define a metric on the
elementary triangles.  The simplest assumption is that the
triangulated surface is composed of flat identical equilateral
triangles. The triangles can be either empty, with weight $T$, or
contain a loop segment making right or left turn at $\pi/6$:
  \bee \begin{tikzpicture}[scale=0.7, tri/.style={fill=yellow!20,
  draw=black, line width=1.0pt}, loop/.style={red, line width=1.5pt,
  decoration={markings, mark=at position 0.49 with { \arrow[red, line
  width=1.5pt]{Stealth[length=7pt]}; } }, postaction=decorate } ]
% Left triangle: empty (weight T)
\draw[tri] (0,0) -- (2,0) -- (1,{sqrt(3)}) -- cycle;
\node[below] at (1,0) {$T$};
% Middle triangle: right turn (weight w)
\begin{scope}[xshift=3.5cm]
  \draw[tri] (0,0) -- (2,0) -- (1,{sqrt(3)}) -- cycle;
  \draw[loop] (0.5,{sqrt(3)/2}) -- (1,0.55) -- (1.5,{sqrt(3)/2});
\node[below] at (1,0) {$w$};\end{scope};
% Right triangle: left turn (weight w^{-1})
\begin{scope}[xshift=7cm]
  \draw[tri] (0,0) -- (2,0) -- (1,{sqrt(3)}) -- cycle;
  \draw[loop] (1.5,{sqrt(3)/2}) -- (1,0.55) -- (0.5,{sqrt(3)/2});
\node[below] at (1,0) {$w^{-1}$};\end{scope}
\end{tikzpicture}
\la{Choice1} 
\eee
The phase factor for each loop in the expansion \re{loopexp7v} can be
thought of as a holonomy factor for a particle with spin.  Each
left/right turn contributes an elementary holonomy $ \exp(\pm i\pi
\l/2)$, a lattice analogue of the spin connection defined on pairs of
adjacent links.  The phase of the holonomy factor, a discrete analogue
of the Berry phase \cite{10.1098/rspa.1984.0023}, is
\bee
 \hf  \pi \l  [ (\text{left turns})-(\text{right turns})]
= \textstyle
    {3\over 2}    \l  \hat K_\CL  
\la{berryphase}
\eee
where $ \hat K_\CL$ is the total geodesic curvature of the oriented
loop $\CL$.  By the Gauss-Bonnet formula, the geodesic curvature is
related to the integrated Gaussian curvature $\hat R_\CL$ inside the
domain $\CT_\CL$ bounded by the loop,
 \bee
 \la{GBf}
  \hat R_\CL+ 2 \hat K_\CL = 4\pi, \qquad 
   \hat R_\CL\equiv \sum_{i\in\CT_\CL} \hat R_i ,
 \eee
where the Gaussian curvature at the site $i\in \CT$ is related in our
case to the coordination number $C_i$ by
   \bee \hat R_i = {2\pi\over 3}\( 6- C_i\).  \eee
For a sphere, changing the orientation of a loop exchanges the
internal and the external domains and the phase factor changes sign.

This is not the only choice for the geometry of the dynamical
triangulations.  We can use triangles with edges of different length
with the condition that only edges of the same length can be glued
together.  For example, we can choose the empty triangles to be
equilateral with sides of unit length, assume that the triangles
containing segments of loops are isosceles with one angle $\pi/2$ and
two angles $\pi/4$ and long side of unit length:
  \bee \begin{tikzpicture}[scale=0.7, tri/.style={fill=yellow!20,
  draw=black, line width=1.0pt}, loop/.style={red, line width=1.5pt,
  decoration={markings, mark=at position 0.5 with { \arrow[red, line
  width=1.5pt]{Stealth[length=7pt]} } }, postaction=decorate } ]
% Left triangle: equilateral, empty (weight T)
\draw[tri] (0,0) -- (2,0) -- (1,{sqrt(3)}) -- cycle;
\node[below] at (1,0) {$T$};
% Middle triangle: right angle at top, right turn (weight w^{-1})
\begin{scope}[xshift=3.5cm]
  \draw[tri] (0,0) -- (2,0) -- (1,1) -- cycle;
  \draw[loop] (0.5,0.5) -- (1,0.05) -- (1.5,0.5);
 \node[below] at (1,0) {$w$};\end{scope}

% Right triangle: right angle at top, left turn (weight w)
\begin{scope}[xshift=7cm]
  \draw[tri] (0,0) -- (2,0) -- (1,1) -- cycle;
  \draw[loop] (1.5,0.5) -- (1,0.05) -- (0.5,0.5);
 \node[below] at (1,0) {$w^{-1}$};\end{scope}
\end{tikzpicture}
\la{choice2}   \eee 
Now the loop segments make turns at $\pm \pi/2$ and the phase factor
of a loop $\CL$ is $ \l \hat K_\CL $.  The relation \re{GBf} still
holds, but the local curvature now is defined as
\bee
\hat R_i = 4\pi - 2\times(\text{the total angle   at the site } \ i).
\eee
Note that a flat triangulation with the choice \re{Choice1} contains
conical curvature defects with the choice \re{choice2} and vice versa.
Furthermore, if $T=0$, the triangles with the choice \re{choice2} can
be glued pairwise along their long sides to make squares.  This brings
us to the formulation of the six-vertex model on random
quadrangulations
\cite{Kostov:1999qx,ZinnJustin:1999wt,ELVEYPRICE2023105739}.

To summarise, in the loop expansion of the gravitational 7v model, the
oriented loops are weighted by a geometrical phase proportional to the
geodesic curvature of the loop.  The definitions of the geodesic and
the Gaussian curvatures depend on the choice of the geometry of the
triangular tiles the world surface is made of, but for the sum over
surfaces this choice is irrelevant.

\subsection{ The 7-vertex matrix model }
 \la{sect:4}
 
Now let us formulate the matrix model which is the holographic dual to
the gravitational 7v model.  The partition function of the 7v matrix
model is given by the integral
  \bee \CZ _N &= \int d\X d\mathbf{Z} d\mathbf{Z}^\dag \ e^{-{1\over
  \hbar} \CS }, \\
   \CS &= \Tr\( \hf \X^2 + \mathbf{Z} \mathbf{Z} ^\dag - {\textstyle
   {T\over 3}} \X^3 - w \X \mathbf{Z} \mathbf{Z} ^\dag  - w^{-1}\X
   \mathbf{Z}^\dag \mathbf{Z} \).  \la{SLMM}
   \eee
The integration variables $\X$ and $ \mathbf{Z} $ are respectively
Hermitian and complex $N\times N$ matrices with flat integration
measure
	\bee d\X d\mathbf{Z} d\mathbf{Z}^\dag = \prod _{i,j} dX_{ij}
	dZ_{ij} d\bar Z_{ij}.  \eee

The perturbative free energy of the 7v matrix model is a sum over all
connected planar (i.e. with thickened lines) Feynman graphs with
vertices as in fig.  \ref{vertices}.  The topology of the planar
graphs is controlled by the Planck constant $\hbar \sim 1/N$, and the
asymptotic semiclassical expansion of the free energy of the matrix
model is also a topological expansion,
\bee \hbar^2 \log \CZ_N =\sum_{g\ge 0} \hbar ^{2g} \CF_{g}( \kappa,
\l),\quad \kappa = \hbar N, \la{thooftexp} \eee
with $\CF_{g}( \kappa, \l)$ being the  contribution
of  the  planar graphs with $g$ handles.

The gaussian integral over the complex matrix can be done explicitly,
leaving an integral over the hermitian matrix $\X$ which, after
factoring out the volume of the $\mathbf{SU}(N)$ group, reduces to an
$N$-fold integral with respect to the real eigenvalues $x_1,..., x_N$,
 \bee \textstyle \mathcal{\CZ} _N & \sim \int\limits _{-\infty
 }^{\infty}\prod_{j=1}^N {dx _i } \ e^{- {1\over \hbar} \( {1\over 2}
 x_j^2 - {T\over 3} x_j^3\)}\ { \prod_{k < j} (x _k- x _j)^2 \over 
 \prod_{k \ne  j} (1- w x
 _k - w^{-1} x _j )} .  \la{eigvalinx1bis} \eee

Strictly speaking, the integral is divergent because the cubic
potential is not bounded from below.  In addition, the integrand has a
simple pole at
\bee x_i = {1/r} , \ \ \ r= 2 \cos \left( \pi \lambda
/2\right)>0\qquad (i=1,..., N).  \eee
However, if we are interested only in the large $N$ expansion, we can
ignore these divergencies.  With the assumption $ \l\in[0,1]$, for
$T>0$ the position of the pole is between the minimum at $x=0$ and the
maximum at $x=T$ of the external cubic potential.  If the pole is
outside of the equilibrium distribution of the eigenvalues around the
minimum we can, up to exponentially small corrections, restrict the
integration to $x<1/r$.  After a linear change of the variable sending
the pole to the origin, the eigenvalue integral takes the form, up to
a normalisation,
 \bee \textstyle \mathcal{\CZ} _N 
& \sim \int\limits _{-\infty
 }^{0}\prod\limits_{j=1}^N {dx _i } \ e^{- {1\over \hbar}  V(x_j)}\ 
{  \prod\limits_{k< j}  (x _k- x _j )^2. \over  
\prod\limits_{k,j} ( q^{1/2} x _k  - q^{-1/2}  x _j )} 
 , \qquad q= e^{i\pi(1+\l)}, 
 \la{eigvalinxter}
  \eee
where $V(x) $ is a cubic potential.  For $T=0$, this is the eigenvalue
integral for the 6v matrix model \cite{Kostov:1999qx}.\footnote{
Integrals of this type have appeared previously in different contexts:
relativistic membranes \cite{1982PhDT........32H}, dimensional
reduction of SYM \cite{Kazakov:1998ji,Hoppe:1999xg}, Euclidean black
hole \cite{Kazakov:2000pm}, Leigh-Strassler deformations of
supersymmetric gauge theories \cite{Dijkgraaf:fk,Dorey_2002}, and more
recently, with non-polynomial potential, as matrix models for
topological strings on Calabi-Yau threefolds
\cite{MarcosMarino-2015,zakany2019matrix}.  Combinatorial aspects of
the planar graph expansion of the model \re{SLMM} were addressed in
\cite{bousquet2025refined}.}

In view of the comparison with the MQM approach to sine-Liouville
gravity \cite{Alexandrov:2002fh}, we will consider the grand canonical
ensemble
\bee \la{Zgc}   \CZ(\bmu) = \sum_{N=1}^\infty e^{ -N \bmu / \hb }\
\mathcal{\CZ} _N  
 \eee
where the chemical potential $\bmu$ plays the role of the lattice
cosmological constant.  The grand partition function is a Fredholm
determinant
\bee \la{Frdet} \CZ(\bmu ) & = \det\( 1+ K\), 
\eee
with the Fredholm kernel 
 \bee \la{Frkn} K(x_ ,x ')\equiv { e^{ -{ \bmu +V(x ) \over  \hbar} }
 \over \qqm x - \qqp x '} \, , \eee
defined with flat integration measure $dx$ on the negative axis.

To compute the genus-zero free energy $\CF_0$ which is the sphere
partition function of the gravitational 7v model, we take the
thermodynamical limit $\hbar\to 0, N\to\infty $ with $\k= N\hbar $
fixed.  Then the sum in \re{Zgc} is saturated by the saddle-point
 \bee \la{mutok} \p_\kappa \CF_0(\k) = \bmu, \qquad \CF_0(\k) =
 \hbar^2 \log \CZ_{\k/\hbar}.  
 \eee
The grand free energy $\CF(\bmu)= \hb^2 \log \CZ(\bmu) $ is related to
the canonical one by Legendre transformation
\bee \tilde \CF_0(\bmu) = - \bmu \kappa+\CF_0 (\kappa).  \eee

The basic observable, which serves as a building block for most of the
observables, is the $L$-th moment of the matrix $\X$,
\bee W_L= \hbar\< \Tr (\X^L)\>  
 , \eee
where the expectation value of the trace in the grand ensemble
\re{Zgc} is defined as the sum, with a weight $e^{- \bmu N/\hbar}$, of
the unnormalised expectation values in the sectors with $N$
eigenvalues.  The resolvent of the random matrix
\bee \la{resolvent} W (x)= \sum _{L=0}^\infty x^{-L-1}
W_L=\hbar\left\langle \Tr {1\over x-\X}\right\rangle \eee
 represents  the disk amplitude $W(x)$ with marked boundary and
(complexified) boundary cosmological constant $x$.

\subsection{ Collective field  and Virasoro constraint}

The grand partition function \re{Zgc} has the form of a Coulomb gas
and can be formulated in terms of a collective bosonic field to be
with mode expansion
 \bee \la{expmodesinf} \bPhi(x) = \bPhi_0 + \bal_0 \log x -\sum_{n\ne
 0} {x^{-n}\over n}\bal_n , \quad [\bal_n,\bal_m]= n\, \d_{m+n,
 0},\quad [ \bPhi_0, \bal_0 ]=1.  \eee
The Hilbert space is built on the Fock vacua defined by 
\bee
 \<0|\bal_{n}=0 \quad (n\le 1 ), \qquad  \bal_{0}|0\> =0 \quad (n\ge 1),
 \qquad 
 \<0|\bPhi_{0} = \bal_{0}|0\> =0 , 
 \eee
and the two-point function is
\bee \< 0| \bPhi (x) \bPhi(x') |0\> = \log(x-x').  \eee

To generate the pairwise interaction of the eigenvalues, which is that
of dipoles composed of a positive charge at $\qqp x_j$ and a negative
charge at $\qqm x_j$, we must introduce two more bosonic fields,
$\bS(x) $ and $ \bPhi( x)$, related to the principal field $\bPhi(x)$ as
\bee \mathbf{\bS}(x) &={ \bPhi( \qqp x)- \bPhi( \qqm x) \over i}
 \\
	\bPhi(x) &= \bal_0 \log x + {\bO( \qqp x)- \bO( \qqm x)\over i} 
	.
	\la{OmSphi} 
	\eee
	The mode expansion of the two auxiliary fields is, with $\b=
	\pi(1+\l)$,
	 \bee \la{mdexpSOm} \mathbf{\bS}(x) &= \b \bal_0 - \sum_{n\ne 0} 2
	 \sin (\hf n\b) \ {x ^{-n} \over n} \bal_n, \\
\bO(x) &= {1\over \b} \bPhi _0 \log x- \sum_{n\ne 0} {1\over 2 \sin
(\hf n\b) } \ {x ^{-n} \over n} \ \bal_n.  \eee

the correlation function of these two fields is
\bee \la{OmegaS} \< 0|\bO(x) \bS(x')|0\> = \log( x-x').  \eee

The field $\bS(x)$ generates the Coulomb interaction of the
eigenvalues and the dual field $\bO(x)$ generates the external
potential.  With the help of these two fields, we express the grand
partition function \re{Zgc} as a Fock space expectation value
   \bee \la{expv} \CZ(\bmu ) & = \<0| \bU _- \bU _+ |0\>
   \eee
    of the product of  the exponential operators  
 \bee \la{partfOp} \bU _- & = \exp\[ -{1\over \hbar } \( \oint_\infty
 (\bmu + V(x) ) {d \bO(x)\over 2\pi i} \) \]
 , \\
  \bU _+ & = \exp\[ \int _\IR dx\ : e^{- \bS(x)}: \]
  ,
   \eee
 with the normal ordering defined so that the operators $\bal_n$ with
 $n\ge 0$ are on the right.  The partition function is generated by
 expanding the exponential in the second factor and applying the
 operator product expansions
 \bee \la{OPES} \bO(x) :e^{ -\bS(x')}: \ & \sim\ \log( x-x')\, :e^{-
 \bS(x')}: \\
   :e^{- \bS(x)}: \ :e^{- \bS(x')}: \ & \sim\ {(x-x')^2\over ( x-
   \qq^{-1} x')( x- \qq x')} \ :e^{- \bS(x)} e^{ -\bS(x')}: \, , \eee
 With the expectation value of the operator $\mathbf{O} $  defined
 as
\bee \< \mathbf{O} \> = \CZ(\bmu)^{-1} \<0| \bU _- \mathbf{O} \bU _+
|0\>,
\la{ptftau} \eee
the oscillators $\bPhi_0$ and $\bal_0$ represent $\bmu$ and
$\p/\p{\bmu}$:
\bee \la{muddmu} 
 { \bmu } &=\b  \hbar \<\bal_0\>
 ,
 \quad 
   {\p  \CF(\bmu )\over \p \bmu}&=-{\hb \over\b}  \< \bPhi_0
\> 
  . \eee
 where $\CF(\bmu)= \hb^2 \log \CZ(\bmu) $ is the all-genus free energy
 of the matrix model.

 We normalise the expectation values so that in the limit $\hbar\to 0$
 they remain finite: $ S(x)\equiv \hbar \< \bS(x)\>$, etc.  The
 expectation value of $\bS(x)$ consists of a polynomial piece,
 $S_\reg(x) = V(x)+\bmu$, and a piece regular at infinity:
\bee S(x) \equiv \hbar \< \bS(x)\> = S_\reg (x) + O(1/x), \qquad
S_\reg(x)= V(x)+\bmu .  \eee
Similarly for $\bPhi$ and $\bO$, with $ \vp_\reg$ and $\O_\reg$
determined by the defining relations \re{OmSphi}, and for the currents
 \bee \bH(x)= \p \bS(x), \ \ \bY(x)= \p \bPhi(x), \ \ \bJ(x)=\p \bO(x).
 \eee
The basic observable \re{resolvent} is given by the regular at
infinity piece of the expectation value of the current $\bJ(x)$:
\bee \la{resope} W(x)= \hbar \<\bJ(x) \> -J_\reg (x) .  \eee

The observables in the matrix model can be obtained with the help of a
conformal Ward identity for the collective field, often referred to as
the Virasoro constraint, which follows from the OPE
  \bee :\bH(x) ^2: \ :e^{ -\bS(x')}:\ &\sim \p_{x'} \({1\over x-x'} :
  e^{ -\bS(x')}: \)+\text{regular}.  \eee
The total derivative form of the rhs insures that the  operator
$:\bH^2(x):$ commutes, up to exponentially vanishing boundary terms,
with the operator $ \bU _+ $ in \re{expv}.  Hence for any operator
$\mathbf{O}$
\bee \la{SDPhi} \oint\limits_\CC{d x'\over x-x'}\ \< \mathbf{O}
:\bH(x') ^2:\> =0\, , \eee
where the integration contour encloses the real axis and leaves
outside all other singularities of the integrand, including the point
$x$.  The Virasoro constraint \re{SDPhi} is equivalent to the
statement that the integrand is analytic in the vicinity of the real
axis.  The conformal Ward identity \re{SDPhi} can also be formulated
in terms of the current $\bY(x)$, as the statement that for any
operator $\mathbf{O}$,
  \bee \la{ViraY} \< \mathbf{O}\cdot \bY ( \qqp x)\bY(\qqm x)\> \
  \text{is analytic in the vicinity of the real axis}.  \eee
Of course, the Virasoro constraint can also be derived  directly
for the matrix integral as a Dyson-Schwinger equation reflecting
translation invariance of the measure.  The Ward identities \re{SDPhi}
or \re{ViraY} can be used to reconstruct the whole genus expansion of
the free energy.  Here we are focusing on the observables on the disk
and on the sphere and will take the semiclassical 
limit $\hbar\to 0$ where the expectation values of traces factorise
and the Virasoro constraint becomes a quadratic functional equation for
a c-function.

\section{Thermodynamical limit  and spectral curve}
\la{sect:Therm}

\subsection{Boundary value problem}

 In the limit $\hbar\to 0$, the fluctuations of the collective field
 are suppressed and the basic observables are determined completely by
 the saddle-point spectral density $\rho(x)$ or, equivalently, by the
 resolvent
 \bee W(x)= \hbar \left\langle \Tr {1\over x- \X}\right\rangle = \int
 _{\IR} {dx' \rho(x')\over x-x'} \eee
 which by definition is analytic in the $x$-plane with a cut on some
 interval $[-\mra,-\mrb]$ with $\mra>\mrb> 0$.  The expectation values
 of the three currents,

 \bee \la{WXH-eqs} J(x)&=\p \O(x)= W(x) + J_\reg(x), \\
 Y(x) & = \p \vp(x) = {\qqp W(\qqp x)-\qqm W( \qqm x)\over i} +
 Y_\reg(x) , \\
  H(x) &= \p S(x)= 2 W(x)- \qq W(\qq x)- \qq^{-1}W( \qq^{-1}x) +
  H_\reg(x), \eee
have respectively one, two and three cuts on rays obtained by rotating
the real axis.
 
As the resolvent $W(x)$ itself does not have simple form even in the
continuum limit, it is more convenient to formulate the observables in
terms of the function $y=Y(x)$ which will be the classical spectral
curve for our problem.  The spectral curve is determined uniquely its
asymptotics at infinity together with the classical Virasoro
constraint \re{ViraY}, which is equivalent to the boundary value
problem
 \bee \la{eqspa} \qqp Y( \qqp e^{\mp i\e} x) = \qqm Y(\qqm e^{\pm i\e}
 x) , \qquad x\in\ [-\mra,-\mrb] .
  \eee
This  condition  for $Y(x)$  is equivalent to  $S(x)$  
being  constant on the interval $ [-\mra,-\mrb] $.
By the second equation \re{muddmu}, the constant is equal to $\bmu$.
 
Once the function $Y(x)$ is known, the classical resolvent $W(x)$ is
completely determined by the relations \re{WXH-eqs}.  To express the
spectral density in terms of $Y(x)$, we write the second relation
\re{WXH-eqs} as
\bee \la{defyJ} W( x) &=i \qqp Y( \qqp x) + e^{i \b } W( \qq x ) \, +
\text{polynomial in $x$}.  \eee
Since $W(x)$ is, by definition, a real analytic function, only the
first term in the r.h.s.~has a cut $[-\mra,-\mrb] \subset \IR$.  The
discontinuity of $W(x)$ across this cut is therefore that of the first
term, and equal to twice its imaginary part,
  \bee \label{spectrd} \rho(x) &\equiv {1\over \pi}\Im W(x)
= {1\over   \pi}\ \Re\[ \qqp Y(   \qqp )\]
 \ \text{for}\ \  x\in[-\mra,-\mrb].\eee

The spectral curve of our problem defines an infinitely foliated
Riemann surface.  By introducing an elliptic uniformisation map to
resolve the four branch points, eq.  \re{eqspa} can be transformed
into periodicity along the $A$-cycle and quasi-periodicity along the
$B$-cycle of the parametrisation torus.  In the $x$-plane, the
$A$-cycle is represented by a closed contour enclosing the upper cut
while the $B$-cycle is represented by an open contour $\CC_B$
connecting the upper edge of the lower cut to the lower edge of the
upper cut.  Then we we can write eq.  \re{muddmu} in the
thermodynamical limit as
  \bee \la{maincrcls} {\bmu }= {1\over i}\int_{\CC_B} {Ydx}, \quad \p
  _{\bmu} \CF_0 =-{1\over 2\pi }\oint _{\CC_A} {Y dx } \, .  \eee

\subsection{Phase diagram and  scaling limit}

The gravitational 7v model is a deformation of the gravitational
$O(2)$ loop model and should have qualitatively the same phase diagram
characterised by dense, dilute and massive phases of the loop gas.  To
describe the critical phases let us return to the canonical ensemble
(fixed $\k$) where the phase diagram is drawn in the $(\k,T)$-plane.

When $\kappa$ increases with $T$ fixed, at some critical value
$\k=\k_c(T)$ the partition sum diverges because of the volume of the
planar graphs.  This is the phase of pure gravity.  On the other hand,
if we keep $\k$ constant and decrease the temperature, below some
$T=T_c(\kappa)$ the effective tension of the loops will become
negative and the planar graph will be densely filled with loops.  This
is the critical phase of dense loops with effective bare cosmological
constant $\k_1=\k/T$.  The two critical lines join at the tricritical
point $(\k_*, T_*)$ where the length of the loops diverges but the
volume not occupied by loops also diverges.  At this point a third
critical phase emerges, the phase of dilute loops.

In the grand ensemble, the critical point is at $(\bmu_*, T_*)$ with
$\bmu_*$ related to $\k_*$ by \re{mutok}.  The continuum limit
concerns the blown-up vicinity of the tricritical point in the double
scaling limit parametrised by the renormalised coupling constants
\bee x\to \e x, \ \ \ \bmu _*- \bmu \to \e^{ 2/(1-\l)} \, 2\pi \mu ,
\quad T -T_* \to \e^ {4\l/(1-\l^2 )} t, \la{sccoups} \eee
where $\e $ is a small elementary length and where we anticipated the
critical indices obtained from the analytic solution we are going to
derive below.  The properly normalised observables in the limit $\e\to
0$ will be given by scaling functions of the dimensionless
combinations $\hat t \sim t/\mu^{ 2\l /(1+\l)}$ and $\hat x \sim
x/\mu^{ (1-\l)/2}$.  The dense critical phase is achieved at $t\to
-\infty$ and is characterised by another set of exponents.

 \subsection{ Uniformisation of the boundary value problem in the
 scaling  limit}

The exact solution of the boundary value problem will be reported in a
future publication \cite{Andre-Ivan-Mateus}.  Here we  focus on
the continuum limit when the parametrisation torus degenerates into a
cylinder, with the large cycle defining a UV cutoff.

\begin{figure}[h!]
        \centering
        \vskip 0.5cm
 \hspace{ 1.5cm}\begin{minipage}[t]{0.4\linewidth}
            \centering
\hskip - 2cm \includegraphics[width=7.0 cm]{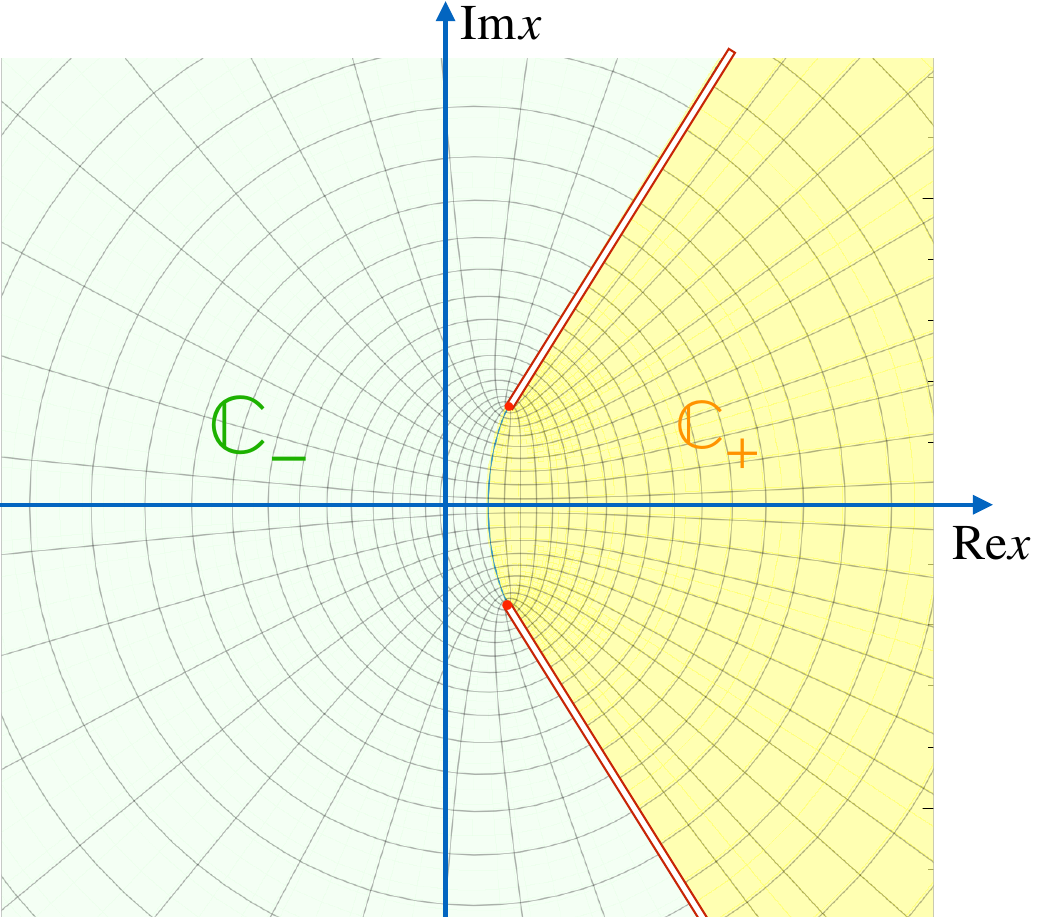} \end{minipage}
 \hskip  0cm   
\begin{minipage}[t]{0.4\linewidth}
            \centering
           \includegraphics[width=6.5cm]{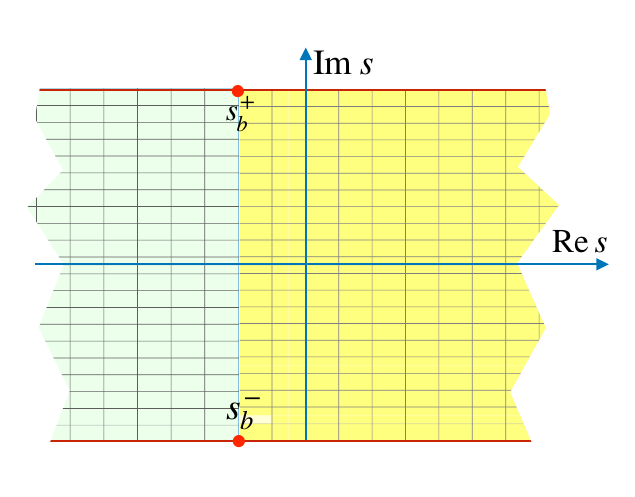}
 \end{minipage}
   \caption{ \small The domains $\IC_+$ and $\IC_-$ of the principal
   sheet of the Riemann surface of $Y(x)$ (left) and in the principal
   strip of the $s$-parametrisation plane (right)}
     \label{TwoDom} 
     \end{figure}

The  scaling  limit is achieved by taking $\e  \sim  \mrb/\mra\to 0$ and
simultaneously rescaling $x$ so that the right branch point stays at a
finite distance $\bb$ from the origin while the left branch point
escapes to  $-\infty$.  From now on,  $x$ denotes the
rescaled variable $x$.  The analytic properties of the function
$y=Y(x)$ change accordingly.  In the  scaling  limit, the Riemann
surface of $Y(x)$ has two semi-infinite cuts in the main sheet
starting at $q^{\pm 1} \bb$ and ending at infinity in the lower/upper
semi-plane $x$ (fig.  \ref{TwoDom}, left).  Such analytic structure
results in two distinct points at infinity, $\Re x\to +\infty$ and $\Re
x\to -\infty$.  One can imagine the Riemann surface as an infinite
cover of a sphere with singularities at the north and at the south
poles.  At the two singular points, the function $Y(x)$ has two
different expansions in fractional powers of $1/x$, convergent
respectively in the domains $\IC_ +$ and $ \IC_-$ delimited by the two
cuts and presented in fig.  \ref{TwoDom} in yellow and green.

The two cuts are resolved by a uniformisation map $ s\to x$ such that
the main sheet of the Riemann surface of $Y(x)$ is parametrised by a
strip of width $\pi$ in the $s$-plane (fig.  \ref{TwoDom}, right)
  \bee \la{paramxs} x(s) &= 2 M \ e^{-\l s} \sinh s , \qquad s\in \IR
  \times [- \hf i \pi, \hf i \pi].  \eee
The whole Riemann surface is parametrised by extending the map
\re{paramxs} to $s\in\IC$.  The upper and the lower cut are the images
of the upper and the lower edge of the parametrisation strip and the
pre-images of the two singular points are at $\Re s\to \pm \infty$.

The parameter $M$ determines the positions of the branch points where
$x'(s) =0$.  Besides the two simple branch points in the main sheet,
there is an infinite number of simple branch points in the lower
sheets whose parameters form a vertical array $ \sb+2\pi i(n+1/2)$
with $n\in \IZ$ and
\bee \textstyle \sb = \log b <0 \qquad (b^2 = {1-\l\over 1+\l} ) .
\eee
The two branch points $\xb^\pm$ in the first sheet are
  \be \xb ^\pm = x(\sb \pm i\pi/2 ) =- \qq^{\pm 1/2} \bb, \ee
 with 
 \be \la{MbM} \bb = \left( 1+b^{-2}\right) b^{\frac{2}{1+b^2 }} M. \ee
The domain $\IC_+$ is parametrised by the semi-infinite strip $|\Im
s|<\pi/2, \ \Re s > s_b $ and the complementary domain $\IC_-$ is
parametrised by the half-strip $|\Im s|<\pi/2, \ \Re s < \sb $, as
depicted in fig.  \ref{TwoDom}, right.
 
The uniformisation map \re{paramxs} transforms the boundary value
problem into a monodromy problem.  The sewing condition \re{eqspa} for
$Y(x)$ can be formulated as a quasi-periodicity condition for the
functions $x(s)$ and $y(s)=Y[x(s)]$,
   \bee
\label{periodis} 
x(s \pm i \pi/2 ) &= e^{\pm i \b} x(s\mp i\pi/2), \\
	y (s \pm i \pi/2) &= e^{\mp i \b} y(s\mp i\pi/2).  \eee

The quasiperiodicity condition \re{periodis} has infinitely many
solutions classified by their asymptotics at infinity.  The correct
asymptotics is determined by the external potential but after taking
the scaling limit we do not have access to this information.  In order
to be able to recognise the solution relevant for our problem, let us
first consider the critical point where $\bb\to 0$.

\subsection{ Solution at the critical point $(M=0)$ }

At the critical point, the bridge connecting the two domains $\IC_+$
and $\IC_-$ disappears and the solution must be sought
separately in each of them,
\bee \la{crsl1} \IC_-: &\qquad Y(e^{i \b} x)&=& \ e^{-i\b} Y(x)\qquad\
&\Rightarrow & \quad Y (x) &\sim & \ (-x)^{ {2 n\over 1+\l}-1} , \\
\la{crsl2} \IC_+: &\qquad Y(e^{i (2\pi-\b)} x)&=& \ e^{-i (2\pi-\b)}
Y(x)\qquad &\Rightarrow & \quad Y (x) &\sim & \ \ \ x ^{ {2 n\over
1-\l }-1} .  \eee
The integer $n$ should be positive because the density must vanish at
the origin.  The most general solution for the resolvent is therefore
\bee \la{rescrit} W (x) = \sum _{n\ge 1} t^+_n { x^{{2 n \over
1-\l}-1}\over 2 \sin( \pi n/ b^2)} -\sum _{n\ge 1} t^-_n {(-x)^{{2 n
\over 1+\l}-1} \over 2 \sin( \pi n b^2)}.  \eee
Indeed, from the second relation \re{WXH-eqs} we obtain
  \bee \la{solcp} Y(x)= \begin{cases} -\sum_{n\ge 1} \ t^-_n\,
  (-x)^{{2 n \over 1+\l}-1}\ & \text{ if } x\in \IC_-\, , \\
  \ \ \ \sum_{n\ge 1}\ t^+_n\, \ x ^{{2 n \over 1-\l}-1}\ & \text{ if
  } x\in \IC_+.
\end{cases}
\eee

In the case of a cubic potential before the scaling limit, the solution
must be a combination of the lowest powers in each of the domains,
 \bee \la{solcp1} Y(x)= \begin{cases} - t\, (-x)^{{ 1-\l\over 1+\l} }\
 & \text{ if } x\in \IC_-, \\
	\ \ \ x^{{ 1+\l \over 1-\l} }\ & \text{ if } x\in \IC_+,
\end{cases}
\eee
where we rescaled $x$ to have $t^+_1=1, \ t^-_{1} = t$.  Scaling
solutions of more general form can be obtained by starting with
polynomial potential of higher degree and tuning the coefficients.

 Taking the
discontinuity of the resolvent, one finds for the
critical spectral density
\bee \la{critrob} \rho(x)& = \frac{ |x|^{\frac{1}{b^2}} +t|x|^{b^2}}{2
\pi } , \qquad x<0, \eee
with $b$ defined by \re{parameterss}.  The rhs of \re{solcp1}
determines the large-$x$ asymptotics of the solution in the continuum
limit we are looking for.

\subsection{Solution  in the scaling
limit}
\la{subsection-solsclim}
 
The unique solution of the quasi-periodicity condition \re{periodis}
with asymptotics \re{solcp1} at   $|x|\to\infty $  is
\bee \label{solscalla} x(s) &= 2 M \ e^{-\l s} \sinh s , \\
y(s ) &= e^{(1+\l ) s} \ M ^{ 1+\l \over 1-\l } -t\,  e^{-(1-\l ) s} M ^{
1-\l \over 1+\l }.  \eee
 This solution corresponds to a certain normalisation of the coupling
 constants $\mu$ and $t$ describing  the scaling limit \re{sccoups}.

With a change of the parameter $s\to \o = e^{-2 s/(b+b^{-1})}$, this
solution takes the form \re{solscall} announced in the Introduction.
The function $y=Y(x)$ is determined in the domains $\IC_\pm$ by two
different power series whose positive parts are given in our case by
\re{solcp}.

The spectral curve in the continuum limit matching the general
solution \re{solcp1} has the parametric form
\bee x(s)&= M e^{(1-\l)s} - M^{-(1+\l)s},\\
y(s)&= \sum_{n\ge 1} \( t_n^+ M^{{2 n \over 1-\l}-1}\ e^{(2n+\l )s} -
t^-_n M^{{2 n \over 1+\l}-1}\ e^{-(2n-\l)s} \).  \eee
For a finite number of non-vanishing couplings, this solution
describes the vicinity of a multicritical point of a more general
vertex model, in close analogy with the $O(2)$ model analysed in
\cite{Kostov:1992pn}.  For example, the tricritical point represents
microscopically the vertex model in the presence of vacancies and
dimers with a negative fugacity.

The fact that the solution has different asymptotics at $x\to +\infty$
and $x\to -\infty$ marks an important difference between the
gravitational vertex model and the other solved models of 2D QG. In
the $O(2)$ model, which corresponds to $ \l=0$, the uniformisation map
has the symmetry $s\to - s$.  Because of this symmetry the parameters
of the branch points must be at $\sb^\pm = \pm i\pi/2$ and the two
infinite-order branch points at $\Re s\to\pm \infty$ can be
identified.  This $\IZ_2$ symmetry is present also in the $O(n)$ model
with $|n|<2$ and in all generalised minimal models of 2D gravity, but
not in the 7vM.
  
\subsection{Computation of  the boundary entropy}

The parameter $M=M(\mu, t)$ is related to the position $x=-\bb$ of the
edge of the eigenvalue distribution.  This is the critical value of
the boundary cosmological constant and can be interpreted as the
renormalised boundary entropy of the sine-Liouville gravity on a disk.

To determine the function $M=M(\mu, t)$, we substitute the solution
\re{solscalla} in equations \re{maincrcls}.  We write these equations
in terms of the function
\bee \la{vps} \textstyle \Phi (s) = \int ^s y(s') d x(s') &= c_+
e^{2s}+c_- e^{-2s} + c _3\, s + \text{const} \eee
with
\bee \textstyle c_\pm = \frac{1}{2} (1\mp \lambda ) M^{\frac{2}{1\mp
\lambda }} \ \ \text{and}\ \ \ \ c_3=(1+\lambda ) M^{\frac{2}{1-
\lambda }} -(1-\lambda ) \, t \, M^{\frac{2}{1+\lambda }} .  \eee
In the parametrisation strip, the now infinite $A$-cycle connects the
points $s=-\infty$ with $s=+\infty$, and the $B$-cycle connects the
two edges of the $s$-parametrisation strip.  The second relation
\re{maincrcls} states, in terms of $\Phi(s)\equiv \Phi[x(s)]$, that for
any $s$,
\bee \la{Ssmu} S(s) \equiv { \Phi(s+i\pi/2) - \Phi(s-i\pi/2) \over i } =
 2\pi  {\mu } .  \eee
Only the last term in \re{vps} contributes to the difference and we
arrive at the transcendental (for irrational $\l$) equation
\re{MLpak00} for the boundary entropy $M=M(\mu, t)$,
\bee \la{MLpak0} {\mu } &  = {1+\l\over 2}M^{ 2\over 1-\l}-
{1-\l\over 2} t \, M^{ 2\over 1+\l} .  \eee

   \subsection{Partition function on the sphere and susceptibility}

Now we will find the derivative of the partition function on the
sphere from the first equation \re{maincrcls}.  The contour integral
of $Ydx = d\Phi(x)$ becomes, in the $s$-parametrisation, 
%twice 
the linear
integral along one of the two boundaries of the parametrisation strip,
\bee \la{intdmu} \p_\mu\CF_0 = -  \int\limits _{ \Im s=
\pi/2} {y(s) dx(s) }.  \eee
It is easier to derive an equation for the second derivative
$u=\p_\mu^2 \CF$ often called the susceptibility.  Let us first compute
the derivative $\p_\mu Y(x)$ using the $s$-parametrisation,
 \bee \la{DmuY} \p_\mu Y(x)&\equiv\p_\mu y(s) |_x = \p_\mu M \( {\p y(
 s) \over \p M} \Big|_s- {\p x(s) \over \p M} \Big|_s { y'(s)\over
 x'(s)} \) = {2 \over x'(s)} .  \eee
This relation is in fact quite general and reflects the existence of a
classical one-dimensional Hamiltonian system for which $x=x(s),
y=y(s)$ define a classical phase-space trajectory, with $\Phi(s)$ being
the abbreviated action.  Since $\p_\mu x = 0$, eq.  \re{DmuY} can be
written in the form of Poisson bracket,
\bee \la{Poissonb} \{ x, y\} = 2 , \qquad \text{with}\ \ \ \{ f, g\}
\equiv {\p f\over \p s} {d g\over \p \mu} -{\p g\over \p s} {d f\over
\p \mu} .  \eee
Now we can obtain the derivative $\p_\mu\Phi$ by integrating both sides
of \re{DmuY},
\bee \la{asymvpx} \p_\mu\Phi(s)|_x&=\int \p_\mu [y(s)x'(s)]_xd s=\int
\p_\mu y(s)\big|_x \, x'(s)ds = {2s } \, .  \eee
Since the $A$-cycle becomes infinite in the continuum limit, we
regularise the integral \re{intdmu} by introducing an UV cutoff $\L$.
Retaining only the leading term in the large $x$ expansion, we have
\bee \textstyle u\equiv \p_\mu ^2 \CF_0= -  (\p_\mu \Phi|_{x =\L}
-\p_\mu \Phi|_{ x=-\L} ) 
\approx -{4\over 1-\l^2} \log{\L\over M}.  \la{uofM}
\eee
 Now we can substitute in \re{MLpak0}
 \bee M= e^{ -{1-\l^2\over 4} u} .  \eee
from \re{uofM} to arrive at equation \re{eqstate6v} for the
susceptibility,
 \bee \la{MLpak01} {\mu } &= {1+\l\over 2}e^{-{1+\l\over 2}u
 } - {1-\l\over 2} t \, e^{ - {1-\l\over 2} u }.  \eee
The three critical points of this equation correspond to the expected
three scaling regimes of the sine-Liouville gravity, $t=0$,
$t\to-\infty$ and $t=t_c>0$.  The third critical point occurs when the
derivative of the susceptibility in $\mu$ diverges (a third order
phase transition), or by $\p \mu/\p u|_{t=t_c}=0$.  In the vicinity of
the three critical points, the free energy is given by
 \bee \la{frenex} \CF_0 = \begin{cases} - { 1 \over 1+\l } \mu ^2\log{
 \mu } +\text{regular} & (t\to 0), \\
 -  { 1\over 1-\l } \mu ^2\log\(-  \mu/t  \)+\text{regular} & (t\to -\infty)
   \\
	 -\text{const} \times (t_c-t)^{5/2} +\text{regular} & (t\to t_c -0 ).
\end{cases}
\eee

Since $\mu_B$ scales as $M$, eq.  \re{MLpak0} gives different scaling
of the boundary length with the area both in the dense and in the
dilute phases.  In the dilute critical phase $\ell \sim A^{1-\l\over
2} $ and in the dense phase $\ell \sim A^{1+\l\over 2} $.  Both
scalings are anomalous for $\l>0$ $(b<1)$ in contrast  to  the
gravitational $O(n)$ model, where only the scaling in the dense phase
is anomalous.

 \subsection{UV and IR compactification radii}
 \la{section:Radii}
 
 Here we give arguments for the relation \re{SLcSLd} between the
 compactification circles at $t=0$ and $t\to -\infty$.
 
In the dense critical phase $t\to -\infty$, where the scaling limit is
that of the 6-vertex model coupled to gravity, the comparison with the
Matrix Quantum Mechanics gives for the compactification radius
  \bee
  \la{Rir}
  \RSLd = {1-\l\over 2}.
  \eee
On the other hand, the scaling law we obtained in the dilute critical
phase $t=0$, eq.  \re{sccoups}, matches the scaling \re{scalingSL} in
the UV limit of the sine-Liouville theory if
  \be
  \la{Ruv}
  2 - {1\over R} = {2\l\over 1+\l}\ \ \ \Rightarrow\ \ \ \Rsluv= {1+\l\over 2}.
  \ee
 Eq.  \re{SLcSLd} follows.  
  Comparing \re{Ruv} with \re{R0e0A} and \re{RSLRSG},
  we can identify
  \bee
  \la{relpla}
  p = {1\over \l}.
  \eee

Let us conclude with three observations.  First, the equation of state
\re{MLpak01} has been derived previously in \cite{Kazakov:2000pm} in
the context of MQM perturbed by a condensate of winding modes and its
small $t$ expansion reproduces the correlation functions of the
sine-Liouville perturbation conjectured by Moore \cite{Moore:1992ac}.
The large $t$ expansions is then automatically obtained using the
symmetry of the equation under
	 \bee \mu \to \mu_1= - {\mu\over t},\quad \l\to -\l,
	 \quad t\to - 1/t.  \la{muprim}
	 \eee
The symmetry of the equation is compatible with the phase structure of
the sine-Gordon model and with the expressions \re{Rir} and \re{Ruv}
being related by $\l\to -\l$.

Second, the expressions for the free energy in \re{frenex} are
proportional to the T-dual compactification radii.  Concerning the
free energy, the continuum limit of the gas of oriented loops
describes the fluctuations of the T-dual field $\tilde \Phi $.

 Third, although the lattice parameter $\l $ does not renormalise, the
 relation of the continuous parameter $p$ to the lattice parameter $w=
 e^{i\l/2} $ depends on the lattice realisation.  On the flat
 hexagonal lattice $1/p = 3\l$, on the flat square lattice $1/p =
 2\l$, while on the dynamical lattice $1/p = \l$.

   \subsection{Boundary length distribution}
\la{section:6}

In the scaling limit, the disk partition functions for fixed boundary
constant $x$ and for fixed boundary length $\ell$ are related by
Laplace transformation,
 \bee \label{invLIW} W (x)&=\int_0^\infty d\ell \ \tilde W(\ell )\
 e^{- \ell x}, \qquad \tilde W_{\ell}&= {1\over 2\pi i} \int _{
 \uparrow } dx\ e^{ \ell x} \, W(x) , \eee
where in the last integral the contour goes in the imaginary
direction, to the right of the branch points.  We cannot express the
resolvent in a closed form, but from the second relation \re{WXH-eqs}
which reads, for the inverse Laplace images,
   \bee
\label{YtellJ}
 \tilde Y(\ell) &={1\over i} \( \tilde W(\qqp\ell )- \tilde W(\qqm\ell
 ) \), \eee
we can reconstruct the small $\ell$ expansion of $ \tilde W(\ell)$
from the inverse Laplace image of $Y(x)$.  The integral can be
expressed as a difference of two linear integrals by closing the
contour around the two cuts of $Y(x)$.  After partial integration,
  \bee \label{WintY} \tilde Y(\ell ) & = {1\over 2\pi i} \int _{
  {\uparrow}} d x \, Y(x) \, e^{ \ell x} = -{1\over \ell } \ \Im \int_
  {-\infty}^\infty e^{ \ell x(s+i\pi/2)} d y(s+i\pi/2) .  \eee
Then $\tilde W(\ell)$ is reconstructed from
  \re{YtellJ}  with the help of \re{periodis},
  \bee \label{WintY2} \tilde W(\ell) &= {1\over \ell } \int_ {-\infty
  }^\infty e^{- \ell M \( e^{-(1+\l) s} + e^{ (1-\l) s}\)} d \left(
  M^{1+\l\over 1-\l} e^{ -(1+\l) s}+ t M^{ 1-\l\over 1+\l} e^{ (1-\l)
  s} \right) .  \eee
 The integral \re{WintY} can be expressed in terms of a generalised
 Bessel integral known as the  Kr\"atzel function
 \cite{kratzel1975integral} (see also e.g.
 \cite{math8040526,Kabeer:2024aa}) which we denote by $K^{(b)} _{
 \nu}(z)$ so that for $b=1$ it coincides with the Bessel-K function,
   \bee
 \label{defII} 
K^{(b)} _{ \nu}(2z) &=\hf \int_ {0}^\infty e^{-z(\o^{1/b} +\o^{-b} )}
\o^{ \nu-1}d \o , \qquad \Re z>0.  \eee
The function $K^{(b)} _{ \nu}(z) =K^{(1/b)} _{ -\nu}(z) $ satisfies
the recursion relations
\bee \label{RecrusEqKK} 2\partial _z K ^{(b)}_{\nu} (z)&=- K
^{(b)}_{\nu+1/b} (z) - K ^{(b)}_{\nu-b} (z) , \\
2\nu K ^{(b)}_{\nu} (z) &=b^{-1} z K ^{(b)}_{\nu+1/b} (z) 
-b \, z\, K ^{(b)}_{\nu-b} (z)   .
\eee
and behaves for large positive argument as
\bee \la{semiclas} K ^{(b)}_{ \nu} (2z) &\sim {e^{-z \bb }
\over\sqrt{z}}\ , \eee
where $\bb$ is given by \re{MbM} with $2M=1$.

Now, changing in eq.  \re{WintY} the integration variable to $\o =
e^{-2 s/(b+b^{-1})}$, we write the disk amplitude with marked boundary
as
\bee \label{WIIb} \tilde W(\ell) &= { 1\over \ell } \( {1\over b}
M^{1/ b^{ 2}} \ K^{(1/b)} _{ 1/b}(2M\ell) - t\, b M^{ b^2} K^{(b)} _{
b}(2M\ell) \) .  \ \eee

\section{Relation to Matrix Quantum Mechanics and 2D string theory}
\la{section:7vMMMQM}

\subsection{The classical spectral curve of MQM}

A non-perturbative formulation of the 2D bosonic string theory is
provided by Matrix Quantum Mechanics (MQM)
\cite{KAZAKOV1988171,Brezin:1989ss,ginsparg19902d,gross1990nonperturbativest},
see \cite{1991hep.th....8019K} for a standard review on the MQM
approach.  The idea is to discretise only the gravitational path
integral as the ensemble of planar graphs embedded in a
one-dimensional continuum, the MQM time.  In the scaling limit, the
singlet sector of MQM reduces to the dynamics of free fermions in an
inverted quadratic potential with a stabilising potential wall far
from the top.  In the grand canonical ensemble, the cosmological
constant is related to the Fermi energy.  The maximal value of the
Fermi energy, near which the quantum effects prevail, is
conventionally set to zero.  The thermodynamical limit corresponds to
a Fermi sea filled up to Fermi energy $E_F = -\mu$ with $\mu$
sufficiently far from the maximum so that the non-perturbative effects
can be neglected.  In this regime the dynamics is that of a
one-dimensional classical Hamiltonian system with $H= P^2-X^2$.  The
individual particles follow classical phase-space trajectories $X=x(E,
\t), P=p(E, \t)$.  The profile of the Fermi sea of the unperturbed
system is stationary and follows the trajectory of the particle with
maximal energy $E= E_F$, namely $ \hf(X^2 - P^2)= \mu $, which also
determines the classical spectral curve of the unperturbed MQM. The
spectral density is proportional to the discontinuity of the function
$P(X)$ on the cut $-\infty<X<\sqrt{2\mu}$,
\bee \int \limits_{H(P,X)\le - \mu}dP \wedge dX = \int
_{-\infty}^{-\sqrt{2\mu}}d X\rho(X), \quad \rho(X)=
P(X)=\sqrt{X^2-2\mu} .  \eee

 \medskip

$\bullet$ \ {\it Integrable perturbations by momentum modes.} The MQM
is known to be solvable in a nontrivial, time-dependent background
generated by a finite tachyon source.  Dijkgraaf, Moore and Plesser
\cite{Dijkgraaf:1993aa} showed that when the allowed momenta form a
lattice as in the case of the compactified Euclidean theory, the
perturbation exhibits the Toda integrable structure.  Toda
integrability was exploited to investigate different aspects of the 2D
bosonic string -- winding mode \cite{Kazakov:2000pm} and momentum
mode\cite{Alexandrov:2002fh,Alexandrov:2001cm,
alexandrov2003matrixquantummechanicstwodimensional} condensates as
well as non-perturbative effects due to sine-Liouville instantons
\cite{Alexandrov:2003nn,Alexandrov:2005ac,2024JHEP...01..141A}.  The
explicit construction of Toda flows with a constraint (string
equation) is summarised in \cite{Kostov:2002tk}.

The Toda flows take a simpler form in the chiral representation
\cite{Alexandrov:2002fh,alexandrov2003matrixquantummechanicstwodimensional}
\bee \la{defxpm} \Xpm= {X\pm P\over \sqrt{2}} \eee
where the Hamiltonian $H=\Xp\Xm$ is diagonalised by Fourier
transformation.  The sine-Liouville perturbation is then introduced by
requiring that the wave functions behave at $\Xpm\to\infty$ as
exponentials of $t_{\pm 1} \Xpm ^{1/R}$.
The grand partition function of the theory at finite temperature with
$ \b = 2\pi R$ obeys, as a function of $ t_1, t_{-1}$ and $\mu$, Toda
equation which, together with the string equation, determines the
susceptibility $ \chi= \p_\mu^2 \log \CF_0$ in the thermodynamical
limit through the equation of state
\bee\textstyle \la{eqstSTR} \mu = e^{ - \chi/R} - {1-R\over R}\ t_{1}
t_{-1} e^{ - {1-R\over R} \chi/R} , \qquad R> R_{\text{BKT}}=1/2.
\eee
The restriction $R>1/2$ comes from the requirement that the
perturbation is relevant.  The deformed profile of the Fermi sea is
the classical phase-space trajectory for $E=-\mu$,
\bee \la{speccST} \Xpm= x_\pm(\t),\quad x_\pm(\t) &= e^{ -\chi /2R}
e^{\pm \t} +t_{\mp 1 } e^{ - {1-R\over R} \chi /2R } \ e^{\mp
{1-R\over R}\t}.  \eee
For real $\t$, this is a section of a complex curve where both
variables $X_+$ and $X_-$ are taken real.  The spectral curve depends
only on the product $t=t_1t_{-1}$ because the dependence on $
t_1/t_{-1}$ can be eliminated by rescaling $\Xpm$.  The spectral
density $\rho(X)$ of the perturbed model is given in parametric form
as $\rho\sim x_+(\t)- x_-(\t) $, $X\sim x_+(\t)+x_-(\t)$.

\medskip

$\bullet$ \ {\it Integrable perturbations by winding modes.} The
sine-Liouville perturbation in the T-dual theory is achieved by
imposing $U(N)$-twisted periodicity in time which effectively
introduces a condensate of winding modes
\cite{Gross:1990md,Kazakov:2000pm,Kostov:2001wv}.  The perturbation is
relevant for $R< \tilde R_{\text{BKT}}=2$.  The resulting integral
over the eigenvalues of the twisting matrix, first obtained (for the
unperturbed theory) by Bulatov and Kazakov \cite{BOULATOV_1993},
resembles the eigenvalue integral \re{eigvalinxter} but with the unit
circle as integration contour accompanied by a prescription for
surrounding the poles.

The unitary matrix integral can be mapped to free fermions.  In the
unperturbed theory, the classical fermionic trajectories are given by
a section of the complex curve $ Z\bar Z =\mu $ where the variables
$Z$ and $\bar Z$ are taken to be complex conjugated to each other.

Perturbations by winding modes also satisfy Toda hierarchy
\cite{Kazakov:2000pm} accompanied by a constraint (string equation)
for which the Lax operators were constructed in \cite{Kostov:2001wv}.
The partition function $\CZ = \exp({\CF/\hb^2})$ perturbed by a
condensate of the two lowest winding modes with couplings $t_1$ and
$\bar t_1$ satisfies Toda equation
  \bee {\p\over \p t_{1}}{\p\over \p \bar t_{1}} \CF(\mu)+
  \exp\left({1\over \hb^{2}}[\CF(\mu+i\hb)+\CF(\mu-i\hb) -2
  \CF(\mu)]\right)=0 \eee
  with boundary condition for $t_1\bar t_1=0$
  \bee \la{boundc} \textstyle \CF(\mu)= -\hf R \mu^2\log\mu - \hb^2
  {R+R^{-1}\over 24} \log\mu +\CO(\hb^2).  \eee
In the thermodynamical (dispersionless) limit $\hbar\to 0$, the
solution for the susceptibility $ \chi = \p_\mu^2 \CF $
  \bee \mu = e^{-\chi /R} - (R-1)t_{1}\bar t_{1} e^{(1-R)\chi /R}
  ,\qquad R< 2.  \la{WMMQM} \eee
The classical fermionic phase-space trajectories are given in
parametric form by $Z=z(\o ), \bar Z=\bar z(\o)$, where $\o$ is a
unimodular parameter and
  \bee z&= e^{-\hf \chi}\ \o \ (1+\bar t_1\ e^{{2-R\over 2R}\chi} \
  \o^{-1} )^R\cr \bar z &= e^{-\hf \chi}\ \o^{-1} \ (1+t_1\
  e^{{2-R\over 2R}\chi} \ \o )^R. \eee
This spectral curve appears also in another realisation of MQM as a
normal matrix model \cite{Alexandrov:2003qk}.

\medskip

\subsection{7vMM versus MQM}

$\circ$ {\it Condensate of momentum modes}

\medskip \noindent Let us show that the MQM with momentum-mode
perturbation and the 7vMM have the same spectral curve.  By a linear
change of the parameter, $ s=-\frac{\tau }{\lambda +1}-\frac{\lambda
}{1-\lambda ^2} \log (M)$, the spectral curve of the vertex matrix
model, eq.  \re{solscalla}, takes a form very similar to \re{speccST}
\bee \la{sympar} x(\tau)&= -e^{\tau } M^{\frac{1}{1-\lambda }}
+e^{\frac{1- \lambda }{1+ \lambda }\tau } M^{\frac{1}{1+ \lambda }},
\\
y(\tau)&= e^{\tau } M^{\frac{1}{1-\lambda }} - t\, e^{\frac{1- \lambda
}{1+ \lambda }\tau } M^{\frac{1}{1+ \lambda }}.  \eee
The two curves become identical upon setting
\bee \la{correspobdence7vMQM} \Xp(\t)&= y(\t), \ \Xm(\t)= - x(\t),\
t_{-1} = -1, \\ t_{1} &= - t, \ R = \hf (1+\l), \ 
e^{\chi
}=M^{1+\l\over 1-\l}
.  \eee
Furthermore the equations of state \re{eqstSTR} and \re{eqstate6v}
match if
\bee \la{MQM7vM} \mu|_{_{\text{7vMM}}} =R\, \mu|_{_{\text{MQM}}} , \ \
u R = \chi /R. \eee
Integrating twice, we conclude that in the interval $1/2<R<1$ the free
energies of the 7vMM and the MQM are related by T-duality.

Expanding both sides of \re{MQM7vM} in $t$, we conclude that the
correlation functions of the thermal operator in the loop gas model
and the correlation functions of the operator creating a pair of
tachyons with momenta $P=\pm 1/R$ in MQM are the same.  On the other
hand, the boundary observables in the two theories are obviously
different.  For instance, in  MQM the dimension of the boundary is
half of the dimension of the bulk while in the 7vMM, as we have seen,
 the boundary has anomalous dimension both in the dilute and in
the dense phases.

Another difference is that 7vMM is defined on the interval $\hf< R< 1$
while MQM with a momentum mode condensate is defined for $\hf <
R<\infty$.  In the Minkowskian picture of MQM, both regimes $\hf< R<
1$ and $1<R<\infty$ represent time-dependant backgrounds, with the
correlation functions of exponential fields interpreted in terms of
multiple scattering of massless tachyons off the ``Liouville wall''.
However, the physics in the two regimes is different.  When $ R>1$,
the two chiral sectors of incoming and outgoing tachyons decouple at
infinity and the tachyon scattering is well defined.  We have a
``regular'' background which can be interpreted in terms of multiple
tachyon states reflected from a time-dependent, but always time-like,
Liouville wall.  On the other hand, if $\hf<R<1$, the Liouville wall
becomes at some moment space-like and the outgoing tachyons can no
longer be defined.  In
\cite{balthazar2023timedependentbackgrounds11dimensional}, see also
\cite{Das:2025fkj}, an elaborate physical picture was put forward in
which tachyons emitted by the accelerating Liouville wall get
eventually absorbed by it.

It would be very interesting to understand how this intricate physics
is perceived from the point of view of 7vMM. In the end, the same
branched cut which spoils the standard tachyon scattering
interpretation appears as condensation of eigenvalues of the 6vMM. The
fact that the spectral curve has two different singular points (the
infinite points of the domains $\IC_+$ and $\IC_-$) suggests that the
asymptotic space for the tachyon scattering should be a direct sum of
two sectors and that the scattering matrix is not diagonal.

More specifically, the wave functions, which are the bulk one-point
functions on the disk in the $\ell$-representation \cite{Moore:1991ir,
Moore:1991ag}, can be expressed in the case of the 7vMM in terms of
the generalised Bessel integral \re{defII}.  The scattering off the
Liouville wall of asymptotic states with momenta $\sim i\nu$ is
encoded in the small-$z$ expansion
 \bee \la{Ingen} K ^{(b)}_\nu (2 z)&=-{1\over 2}\sum_{n=0}^\infty \
 C_n(\nu, b) \ \ z^{ b\nu + (1+b^2) n } - {1\over 2}\sum_{n=0}^\infty
 C_n(-\nu, 1/b) \ z^{ - \nu/b + (1+b^{-2})n } \\
 &=\hf b \Gamma (-b \nu ) z^{b \nu } +
 \hf \frac{\Gamma \left(\frac{\nu }{b}\right) z^{-\frac{\nu }{b}}}{b}
 +...  \ ,
\quad
C_n(\nu, b) = (-1)^n
 {b  \Gamma ( -b^2 n-b \nu ) \over n!}\ .
\eee
 The asymptotic space is a direct sum of two sectors related by
 $b\leftrightarrow 1/b$.  The map $b\to 1/b$ exchanges the
 compactification circles at the two endpoints of the gravitational
 massless flow:
 \bee
 \Rsluv= {1+\l\over 2}= { b^{-1}\over b+ b^{-1}}\quad 
 \leftrightarrow \quad { b \over b+ b^{-1}}= {1-\l\over 2}= 
 \Rslir
 \eee
For imaginary $\nu$, the two lowest terms of the expansion represent
incoming and outgoing particles belonging to different sectors of the
asymptotic space.  A first sight, the momentum is not conserved by the
scattering, but actually it is conserved because in the two asymptotic
sectors the boundary length depends differently on the constant
Liouville mode, see the discussion after eq.  \re{muprim}:
\bee
\la{scalingell} \ell &= e^{(1+\l)\phi}=e^{ b^{-1} \phi/(b+b^{-1}} \qquad
&(\text{dense phase}) \\
\ell&=e^{(1-\l)\phi}=e^{ b \phi /(b+b^{-1})} \qquad
&(\text{dilute phase}) .  \eee

 \medskip

\medskip  \noindent
$\circ$ {\it Condensate of winding modes}

\medskip \noindent Now we turn to the case of perturbations by winding
modes.  The equation of state \re{WMMQM} becomes identical to the
equation of state \re{MLpak01} in the 7vMM upon replacing
\bee R \to {2 \over 1+\l} , \quad \chi \to u , \quad t_{1}\bar
t_{1}\to t , \quad \mu\to     {2\mu\over 1+\l} \qquad (1<R<2).  \eee

In \cite{Kazakov:2000pm}, it was suggested that MQM at $R=3/2$
perturbed by strong winding mode condensate might provide a matrix
model for the Euclidean black hole.  The pure sine-Liouville theory
dual to the Euclidean cigar background \cite{FZZconj, 2017} should be
represented by the vicinity of the critical point $t\to -\infty$ of
the equation of state \re{WMMQM}.  This critical point cannot be
achieved by simply setting $\mu=0$ in \re{WMMQM} because by doing that
we miss the logarithm in $\mu$ and obtain only the uninteresting
analytic in $\mu$ part of the free energy on the rhs of \re{frenex}.

Moreover, in \cite{Kazakov:2000pm} it was tacitly assumed that taking
the limit $\mu\to 0$ or equivalently $t\to -\infty$ (after choosing
the correct sign for $t$) does not alter the compactification radius
which is, as follows from our analysis, incorrect.  If we start with
$R=3/2$, the endpoint of the flow will be at $R=3$ as follows from
from the T-dual version of the relation \re{SLcSLd}.  In any case, the
proposal of \cite{Kazakov:2000pm} needs to be carefully reconsidered.

   \section{Discussion}
  
In this paper we proposed a microscopic realisation of the
sine-Liouville gravity based on a ``dilute'' vertex model on a dynamical
lattice which represents a one-parameter generalisation of the
six-vertex model considered earlier.  We extracted the phase diagram
of the theory from the classical spectral curve of the dual large-$N$
matrix model and gave a QFT interpretation of the critical points,
guided  by the phase diagram of the sine-Gordon model on a flat
surface.  In particular, we identified the gravitational equivalent of
the massless flow in the sine-Gordon model with imaginary coupling
constant. We conclude with several remarks.

  \medskip

$\circ$ If the coupling constants $t_{\pm 1}$ in \re{eqstSTR} are
identified (up to a normalisation) with $\tpm$ in the worldsheet
action \re{actionSG}, the worldsheet theory for the point $ t_
+=t_-=0$ is a massless free boson coupled to Liouville gravity.  On
the other hand, by the correspondence \re{correspobdence7vMQM}, the
worldsheet theory for the 7vMM at $t=0$ is defined by the UV action
\re{actionSG} with $\tp=0$ and $\tm \ne 0$.  Concerning the bulk
observables, the two worldsheet theories are identical because of the
charge conservation, but the two theories can have different boundary
observables.  In the unperturbed MQM, the boundary conditions are
imposed independently on the Liouville field and on the free boson and
are hence of FZZT \cite{Fateev:2000ik,Teschner:2001ab} type.  The
boundary cosmological constant is coupled to the area of the world
sheet via the Liouville interaction $\mu_B \, e^{\phi}$.  On the other
hand, in the 7vMM, the disk partition function for $t\to 0$ is given
by the first term on the rhs of \re{WIIb} with $ M\sim \mu
^{(1-\l)/2}$ and certainly does not correspond to a FZZT brane.  The
boundary interaction in the worldsheet theory \re{actionSG} would
necessarily involve both the Liouville and the matter fields.

  \medskip

$\circ$ One can speculate that the boundary term of the sine-Liouville
action also has sine-Liouville form.  According to \re{scalingell},
for $t=0$ the length element is $d\ell \sim e^{(1-R)\phi}$.  This is
the dressing factor for the exponential boundary field $e^{\pm i
R\vp}$.  We therefore conjecture that at $t=0$ the boundary term is of
the form
\bee
\mu_+^B e^{(1-R)\phi} e^{iR\tilde \vp}+\mu_-^B e^{(1-R)\phi }e^{-iR\tilde \vp}, 
\la{bdrytermSL}
\eee
where $\tilde\vp$ is the T-dual field of $\vp$, with $x\sim \mu_B=
\mu^B_+\mu^B_-$.  At the opposite end of the flow, where the Gaussian
field is perturbed by an irrelevant sine-Liouville operator, the
boundary interaction with the correct scaling of the boundary
parameter should again be of the form \re{bdrytermSL}, with $R
\to1-R$.

  \medskip
  
$\circ$ For the known solvable models of 2D gravity, such as the
$O(n)$, Potts and the ADE models on planar graphs, the matter is
represented by clusters, e.g. non-intersecting loops, with fugacities
depending only on the topology of the cluster and not on the geometry
of the lattice.  The 7v model on planar graphs, as well as the 6v
model which is a particular case, belong to a new type of discrete
models of 2D gravity: the gravitational vertex models.  In the vertex
models on dynamical lattices, the weights of the clusters are not
topological but depend on the distribution of the local curvature
throughout the lattice.  When formulated as height model cf.
\re{BweightsSOS}, the 7v model represents a $q$-deformed verson of the
affine $\hat A_1$ height model.  It is quite straightforward to
formulate the $q$-deformed $\hat A \hat D \hat E $ models on dynamical
surfaces and their dual matrix models.

\medskip 

$\circ$ We studied the near-critical behaviour of the vertex model
with a single thermal coupling coupled to the vacancies.  This is the
simplest of an infinite hierarchy of multicritical points.  The
multicritical phases and their microscopic construction for $\l =0$
have been studied in \cite{Kostov:1992pn}.  We expect that the
continuum CFT for the $\hat A_n$ series is again sine-Liouville
gravity, eq.  \re{actionSG}, with $R_\uv= \hf n(1+\l)$ and $R_\ir =
\hf n(1-\l)$.  These models might be interesting as realisations of
the proposal of \cite{Kazakov:2000pm}.

  \medskip

$\circ$ For $\l=0$ ($q=-1$), Alexey Zamolodchikov conjectured
\cite{Zamolodchikov:1994uw} that Fredholm determinants of the type
\re{Frdet} solve the ``TBA-like" integral equations which appear in
certain $\mathcal{N} = 2$ supersymmetric theories
\cite{Cecotti:1992qh} (see also \cite{Cecotti:1991me}).  This
Zamolodchikov conjecture was later proved by Tracy and Widom 
\cite{Tracy_1996}.  The integral equations are better suited
for numerical analysis than the differential equations from Toda
hierarchy and it would be interesting whether they can be formulated
for general $\l$.  The case $\l=1/6$ was considered in
\cite{Okuyama:2015pzt}.

\acknowledgments

I am grateful to Andre Alves Lima and Mateus da Silva Junca for
numerous discussions, and Galen Sotkov for his interest and some
important remarks.  I thank Sergey Alexandrov, Vladimir Kazakov, David
Kutasov and Valentina Petkova for providing useful comments on the
draft version of the manuscript.

%
%%
   % \bibliography{/Users/vani/Documents/PAPERS/ABib-good.bib}
%%%%  
 % \bibliographystyle{JHEP.bst}
%%%

\providecommand{\href}[2]{#2}\begingroup\raggedright\endgroup

\end{document}